\documentclass[aps,prd,nofootinbib,preprintnumbers,amsmath,amssymb,twocolumn,superscriptaddress]{revtex4-2}

\usepackage{amssymb}
\usepackage{graphicx}
\usepackage{amsmath}
\usepackage{hyperref}
\usepackage{multirow}
\usepackage{booktabs}
\usepackage{setspace}
\usepackage{verbatim}
\usepackage{float}
\usepackage{color}
\usepackage{ulem}
\usepackage{arydshln}
\usepackage[utf8]{inputenc}
\usepackage[table,xcdraw]{xcolor}
\usepackage{lipsum}
\usepackage{mathtools}

\graphicspath{{./figures/}}

\newcommand{\blue}{\textcolor{blue}}

\newcommand{\md}{\mathrm{d}}
\newcommand{\transp}{\mathsf{T}}
\newcommand{\gagg}{g_{a\gamma\gamma}}
\allowdisplaybreaks

\newcommand{\JR}[1]{{\textcolor{purple}{[Jing: #1]}}}

\begin{document}

\title{The MeerKAT Thousand-Pulsar Polarisation Array II: \\ Searches for Ultralight Axion-Like Dark Matter} 

\author{Zi-Yan Yuwen}
\affiliation{Department of Physics, Stellenbosch University, Matieland 7602, South Africa}
\affiliation{Asia Pacific Center for Theoretical Physics (APCTP), Pohang 37673, Korea}
\affiliation{Institute of Theoretical Physics, Chinese Academy of Sciences (CAS), Beijing 100190, China}

\author{Michael Sarkis}
\affiliation{Department of Physics, Stellenbosch University, Matieland 7602, South Africa}

\author{Yin-Zhe Ma}
\email{Corresponding author: Y.-Z. Ma, mayinzhe@sun.ac.za}
\affiliation{Department of Physics, Stellenbosch University, Matieland 7602, South Africa}
\author{Tao Liu}
\affiliation{Department of Physics and Jockey Club Institute for Advanced Study, The Hong Kong University of Science and Technology, Hong Kong S.A.R., China}
\author{Jing Ren}
\affiliation{Institute of High Energy Physics, Chinese Academy of Sciences, Beijing 100049, China}
\affiliation{Center for High Energy Physics, Peking University, Beijing 100871, China}
\author{Patrick Weltevrede}
\affiliation{Jodrell Bank Centre for Astrophysics, Department of Physics and Astronomy, University of Manchester, Manchester M13 9PL, UK}
\author{Xiao Xue}
\affiliation{Institut de Física d’Altes Energies (IFAE), The Barcelona Institute of Science and Technology, Campus UAB, 08193 Bellaterra (Barcelona), Spain}

\begin{abstract}
    We construct Pulsar Polarisation Arrays (PPA), using regular pulsars monitored in MeerKAT's Thousand Pulsar Array (TPA) Programme, to search for Axion-like Dark Matter (ALDM) within Milky Way. 
Specifically, from a catalogue of 1237 regular pulsars, we select the 50 ones with the highest signal-to-noise ratio and set upper limits on the ALDM Chern–Simons coupling. 
    We find no signals with statistical significance over the mass range of $[10^{-23},10^{-20}]~\mathrm{eV}$ in the six-year MeerKAT's data. By combining the high-quality TPA pulsars and the accurate ionospheric subtraction of \texttt{spinifex}, we establish the most sensitive upper limits to the date on the ALDM Chern-Simons coupling, namely $\lesssim 10^{-14} - 3\times 10^{-13}\,$GeV$^{-1}$, for the mass range of $[10^{-23},10^{-21}]~\mathrm{eV}$ except at $m_a \sim 1.3 \times 10^{-22}\,$eV. This study underscores the great potential of constructing regular-pulsar PPAs for scientific tasks.  
\end{abstract}
\maketitle

\textit{\textbf{Introduction.}---}
There is multiple observational evidence for the existence of Dark Matter (DM) in the universe, such as galactic rotation curves, galaxy cluster dynamics, CMB power spectra, and large-scale structure formation scenarios.~\cite{cirelliDarkMatter2024}.
Nevertheless, the microscopic nature of DM remains unknown, leading to a variety of particle candidates. Among these, ultralight DM candidates such as axion-like particles, dark photons, and massive gravitons have received significant attention. Due to their low masses, ultralight DM may have a high occupation number in galactic environments, forming a Bose-Einstein Condensate within the halo that exhibits wave-like behavior on macroscopic scales. This can lead to the formation of cored structures in the DM halo near the galactic center. In the scenario often referred to as ``Fuzzy'' DM~\cite{Hu:2000ke,hui2017ultralight}, where the particle mass is around $10^{-21} - 10^{-22}\,$eV, the characteristic scales, determined by the de Broglie wavelength, can extend to hundreds of parsecs. This scenario has been proposed to address small-scale structure issues in astronomy, such as the ``core–cusp'' problem.

Ultralight DM can be probed gravitationally through observations such as the Lyman‑$\alpha$ forest~\cite{Kobayashi:2017jcf,Armengaud:2017nkf}, dwarf‑galaxy dynamics~\cite{Marsh:2018zyw,Hertzberg:2022vhk,Teodori:2025rul,Caputo:2026wmy}, astrometry~\cite{Guo:2019qgs,Kim:2024xcr,Yu:2026ebx}, and pulsar timing~\cite{EuropeanPulsarTimingArray:2023egv,Porayko2024,PPTA:2024mgh}. Although constraints have been reported, these measurements are subject to various astrophysical or theoretical uncertainties, highlighting the value of developing new probes. A complementary approach is to search for their non‑gravitational interactions with Standard Model particles. Below, we focus on ultralight axion-like dark matter (ALDM) as a representative candidate to illustrate the power of such probes.


The ALDM halo can interact with the electromagnetic (EM) field through the Chern-Simons interaction $\sim \frac{1}{2} \gagg aF_{\mu\nu}\tilde{F}^{\mu\nu}$, where $\gagg$ is the coupling, $a$  the ALDM field, and $F_{\mu\nu}$ the EM field strength along with its Hodge dual $\tilde{F}^{\mu\nu}$.  
Within the halo, this interaction splits the dispersion relation between the two EM circular polarisation modes~\cite{carrollLimitsLorentzParityviolating1990}, resulting in a position angle (PA) rotation $\Delta\mathrm{PA} \simeq g_{a\gamma\gamma}\left(a(\mathbf{x}_\mathrm{p},t_\mathrm{p}) - a(\mathbf{x}_\mathrm{e},t_{\rm e})\right)$~\cite{harariEffectsNambuGoldstoneBoson1992}, generally known as cosmological birefringence (CB), for linearly polarized light. This effect is fully determined by the ALDM field profile at the positions of light emission and reception (denoted as ``p'' (pulsar) and ``e'' (Earth), respectively), thereby following some specific oscillating pattern over spacetime.    
%
In the past, various light sources have been suggested to search for this effect, such as CMB~\cite{Lue1999,Feng2006,fedderkeAxionDarkMatter2019}
and the pulsar light~\cite{Liu:2019brz} 
and even the light in terrestrial cavity~\cite{liuSearchingAxionDark2019}. 

Pulsar Polarization Array (PPA) is an astronomical method recently proposed~\cite{Liu:2021zlt} to explore fundamental physics by cross-correlating pulsar polarization data. Since the ultralight ALDM can couple with pulsar polarization data through the CB effect, in a galactic-scale wave pattern, the PPA is especially suited for its detection~\cite{Liu:2021zlt}.
Recently, the Parkes Pulsar Timing Array (PPTA) collaboration built the Parkes PPA (PPPA) using polarization data of millisecond pulsars (MSPs) from its third data release and carried out the first PPA search for ALDM, yielding the world-leading constraints on the Chern-Simons coupling for the mass range of ``Fuzzy'' DM~\cite{PPTA:2024mgh}~\footnote{Note, while the European PTA collaboration performed a parallel analysis using polarization data from multiple MSPs in~\cite{Porayko2024}, 
its analysis did not take into account the pulsar cross correlations, which are a core element of the PPA.}. Simultaneously, the pulsar cross-correlation signatures for the ultralight ALDM have been applied to its PTA detections~\cite{Luu:2023rgg}, which had been overlooked in previous studies~\cite{khmelnitskyPulsarTimingSignal2014,poraykoConstraintsUltralightScalar2014,poraykoParkesPulsarTiming2018,afzalNANOGrav15Yr2023,EuropeanPulsarTimingArray:2023egv}, and further to a synergy of PPA and PTA searches~\cite{Li:2025xlr}.
Despite these achievements, future PPA analyses could be limited by the population of MSPs and their distributions within the Galaxy. As proposed earlier~\cite{Liu:2021zlt}, a PPA comprising a large and widely distributed pulsar sample would be highly valuable for examining spatial correlations of the ALDM signals in a broad mass range. Regular pulsars, which constitute about 85–90\% of the known pulsar population, offer precisely this possibility. Unlike timing - where the extreme rotational stability of MSPs is crucial - polarization profile stability differs much less between regular pulsars and MSPs; in fact, the two groups exhibit comparable median luminosities and similar polarization‑profile characteristics~\cite{Karastergiou:2024wkk}. Consequently, with roughly an order‑of‑magnitude larger population and a much wider spatial distribution across the Galaxy, regular pulsars have the potential for PPA to access smaller spatial correlation scales and to sample the correlation pattern more precisely. This would not only improve statistical sensitivity, but also sharpen the PPA's ability to recognize the nature of candidate signals.



\begin{figure}
    \centering
    \includegraphics[width=0.98\linewidth]{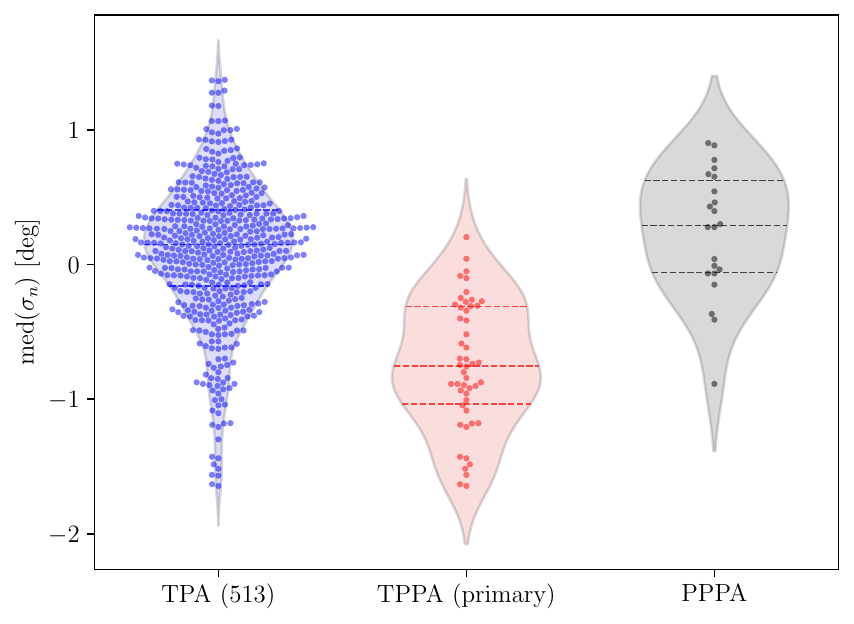}
    \caption{Median of the PA observational errors ($\sigma_n$) for the 513 selected TPA pulsars, the 50 primary-TPPA pulsars and the 22 PPPA MSPs. Here the subscript ``$n$'' denotes the $n$-th epoch in time series of each pulsar. 
    The inner dashed lines in each shaded region denote the median and the interquartile range ($25-75$ percentiles) of the med($\sigma_n$) distributions. }
    \label{fig: sigma_n compare}
\end{figure}

In this \textit{Letter}, we will construct the first PPA of regular pulsars, using the data from the MeerKAT’s Thousand-Pulsar-Array (TPA) Programme~\cite{Johnston:2020qxo}. This programme started off as part of the larger MeerKAT Key Science Programme MeerTime~\cite{bailesMeerTimeMeerKATKey2018}. The statistical polarization properties for these regular pulsars were studied in~\cite{Sarkis:2025}. We then will apply the Thousand-Pulsar Polarisation Array (TPPA) to detect ultralight ALDM, and present its first constraints on the ALDM Chern-Simons coupling.    
\textit{\textbf{TPPA Construction.}---} The TPA catalogue includes 1237 regular pulsars in total. Their $\Delta \mathrm{PA}$ time series are defined by computing the difference between the observed PA and the averaged PA profile from derotation of the Stokes parameters~\cite{Sarkis:2025}, with a maximal timespan of six years. In the preparation, the data are cleaned, by removing individual outliers which might be caused by systematics such as radio frequency interference. By excluding those with valid data points below 20, we have 513 pulsars left.  
As a precursor to the next‑generation radio telescope - Square Kilometre Array, the MeerKAT's instrumentation is more advanced than that of the Parkes telescope. Benefitting from it, this large population of pulsars demonstrate a median of med($\sigma_n$) smaller than that of the PPPA MSPs, as demonstrated in Fig.~\ref{fig: sigma_n compare}, with a tail extended downward to $\sim \mathcal O(10^{-2})\,$deg. This outcome allows us to focus on high-quality pulsars, minimizing unexpected impacts from the low-quality group. Thus, we construct the primary TPPA using the $50$ pulsars with the highest median signal-to-noise ratio (SNR) of Stokes $I$ parameter of all observation epochs (referred to as ``SNR'' group), with the median of their med$(\sigma_n)$ distribution about one order of magnitude smaller than that of the PPPA MSPs~\cite{PPTA:2024mgh}. 
This strongly motivates the exploration of PPA science with the TPA data.

The regular pulsars in the TPPA are widely distributed in the Galaxy, with their space separations and distance to the Earth ranging from hundreds to thousands of parsecs. Different from the PPPA case, this feature offers a flexible basis for selecting different pulsar subgroups, enabling a tailored probe of spatial correlations in ALDM‑induced PA rotations. 
Here we construct two auxiliary TPPAs by picking up $10$ pulsars for each from the 50 pulsars of the primary TPPA, referred to as ``Compact (COM)'' and ``Extended (EXT)'' respectively, to assist the data investigation. The pulsars of the COM TPPA have a distance below $1\,$kpc universally. Their spatial distribution is relatively compact, with the mean  inter-pulsar and pulsar-Earth separation $\sim 0.66 \, \mathrm{kpc}$. As a reference, the mean separation for the PPPA MSPs in~\cite{PPTA:2024mgh} is $1.45 \, \mathrm{kpc}$. 
The EXT group, on the other hand, contains pulsars that are relatively distant and sparse,
all having a distance comparable to or above $2\,\mathrm{kpc}$ to the Earth. The mean inter-pulsar and pulsar–Earth separation is $\sim 4.98 \, \mathrm{kpc}$. %
Consequently, the COM TPPA is good at probing the signal spatial correlations in the relatively high‑mass regime, whereas the EXT TPPA could benefit from enhanced signal strength for those pulsars that lie closer to the Galactic center, where the halo is relatively dense.

See Supplemental Material (SM) at Sec.~\ref{sec: data prepare} and Ref.~\cite{Sarkis:2025} for additional details about TPPA construction.


\textit{\textbf{Analysis Setup.}---} The ALDM halo can be modelled as a stochastic superposition of particle waves, due to its extremely low particle mass and high occupation number, as well as its random origin from primordial density fluctuations ~\cite{khmelnitskyPulsarTimingSignal2014,huiWaveDarkMatter2021}. In this work, we will take a Bayesian method to search for the ultralight ALDM in TPPA data, where the signal signature is  statistically manifested as two-point correlation functions of pulsar time series in the likelihood covariance matrix $C^a$. With the stochastic modelling, the two-point correlation functions between pulsars $p$ and $q$ are given by~\cite{Liu:2021zlt}
\begin{align}\label{eq: signal_covariance_main}
    C^a_{p,n;q,m} &= S_a^2 \bigg\{ \cos\left[ m_a(t_{p,n}-t_{q,m}) \right] \\
    & \quad~ + \sqrt{\rho_p \rho_q/\rho_\mathrm{e}^2} \cos\left[ m_a(t'_{p,n}-t'_{q,m}) \right] \frac{\sin y_{pq}}{y_{pq}} 
    \nonumber\\
    & \quad~ - \sqrt{\rho_p / \rho_\mathrm{e}} \cos\left[ m_a(t'_{p,n}-t_{q,m}) \right] \frac{\sin y_{\mathrm{e}p}}{y_{\mathrm{e}p}} \nonumber\\
    & \quad~ - \sqrt{\rho_q / \rho_\mathrm{e}} \cos\left[ m_a(t_{p,n}-t'_{q,m}) \right] \frac{\sin y_{\mathrm{e}q}}{y_{\mathrm{e}q}} \bigg\}~, \nonumber
\end{align}
where $S_a \equiv g_{a\gamma\gamma}\sqrt{\rho_\mathrm{e}}/m_a$ is the characteristic magnitude of signal, and $t_{p,n}$ and $t'_{p,n}$ the reception and emission time of pulsar light for the $n$-th observation epoch. $y_{ij}=|\mathbf{x}_i-\mathbf{x}_j|/l_{\rm c}$ is dimensionless, denoting the inter-pulsar and pulsar-Earth separations.  $l_{\rm c}=1/(m_a v_0)$ is the coherence length of ALDM with the virial velocity $v_0 \sim 10^{-3}$. 
As the TPPA pulsars are distributed over a range of several thousands of parsecs, we take the best-fit Einasto halo profile inferred from recent photometric data~\cite{ouDarkMatterProfile2024}.
This covariance matrix encodes the full signal correlations. Its diagonal ($p=q$) and off-diagonal blocks ($p\neq q$) represent the ALDM-induced pulsar auto-correlations and cross-correlations, respectively.
The two auxiliary TPPAs behave differently: the COM group, with generally smaller $y_{ij}$, is sensitive to spatial correlations over a larger mass range, while the EXT group contains pulsars whose local ALDM density $\rho_p$ can reach 3-4 times that of $\rho_e$, implying that the pulsar–pulsar term dominates at least in the auto‑correlation.


The noise modelling for this analysis mainly follows the method developed for PTA and PPA studies~\cite{Lentati:2013rla,PPTA:2024mgh}. In addition to observational errors and temporally correlated RM variances of random nature, we include a polynomial of up to second order~\cite{PPTA:2024mgh} 
to account for slow variations in the properties of the Ionized Interstellar Medium (IISM) and the resulting Faraday Rotation (FR), along each pulsar's line of sight. 


One challenge for noise modelling arises from evaluating temporal variance of the ionosphere-induced FR, where multiple effects and time scales could be engaged through plasma profile and geomagnetic field.
As shown in~\cite{PPTA:2024mgh}, 
the use of the recently developed package \texttt{ionFR}~\cite{Sotomayor-Beltran:2013vma} to subtract these effects may leave unmodelled ingredients in the PA residual time series.  
To better assess the uncertainties associated with ionospheric modelling,  
we thus employ an additional public package \texttt{spinifex}~\cite{mevius_2025_15000430} for comparison. 
We present the noise modelling for single pulsars  in~\cite{Sarkis:2025}, which shows that the ionospheric subtraction with \texttt{spinifex} in general results in a significantly better fit of the noise model to data. Below we will compare the performance of these two ionospheric packages in the ALDM search, illustrating how an improved ionospheric model can enhance sensitivity  (See SM 
for additional validations of our noise modelling).


\begin{figure}
    \centering
    \includegraphics[width=\columnwidth]{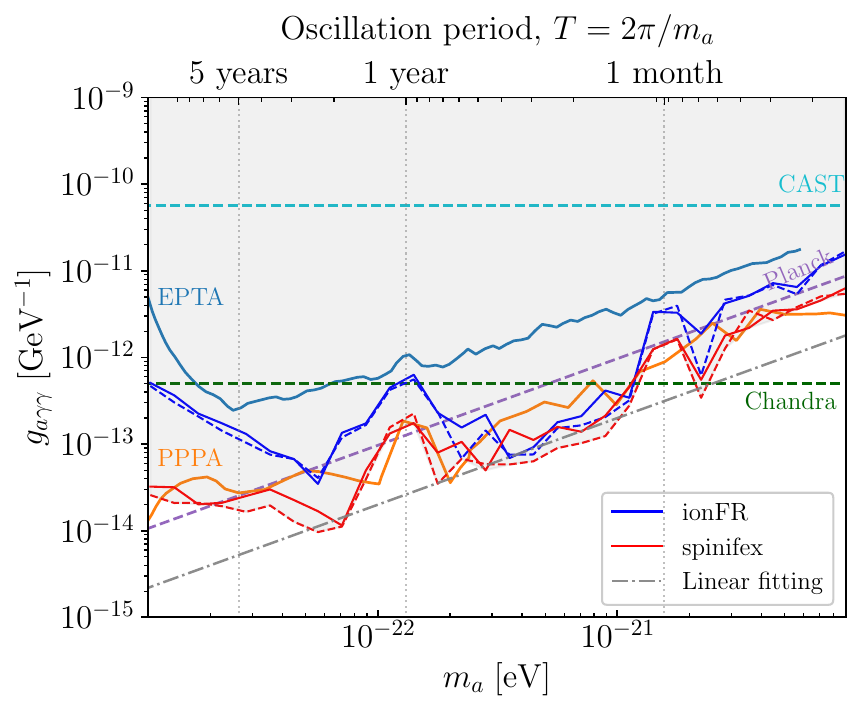}
    \caption{95\% upper limits on the ALDM Chern-Simons coupling $g_{a\gamma\gamma}$ as a function of $m_a$. 
    These limits are obtained with the \texttt{ionFR} (blue) and \texttt{spinifex} (red) packages applied for the ionospheric subtraction. The solid and dashed lines correspond to the constraints for the signal models with the full-correlation across the TPPA pulsars, and with the auto-correlation only for individual TPPA pulsars, respectively.
    The linear fitting function of $g_{a\gamma\gamma}\sim 10^{-13.3} \left(\frac{m_a}{10^{-22}\mathrm{eV}}\right)~\mathrm{GeV}^{-1}$ to the TPPA limits in the high mass regime is presented in dot-dashed grey. As a reference, the limits from the PLANCK CMB polarization measurements~\cite{fedderkeAxionDarkMatter2019},the Chandra Transmission Grating Spectroscopy~\cite{Reynes:2021bpe}, the CAST helioscope at CERN~\cite{castcollaborationNewCASTLimit2017}, and the European Pulsar Timing Array (EPTA)~\cite{Porayko2024} and the PPPA~\cite{PPTA:2024mgh} are also shown. The vertical dotted lines represent characteristic time scales of 5 years, 1 year and 1 month. } 
    \label{fig:gagg_main_main}
\end{figure}

\textit{\textbf{Analysis Results.}---} To generate the constraints on the ALDM Chern-Simons coupling, we take a Bayesian analysis using Markov-Chain Monte-Carlo (MCMC) sampling chains, where the intrinsic parameters of each pulsar are assumed to be free.   
The 95\% upper limits are then extracted from the credible interval of the $S_a$ posterior distribution for a given ALDM mass, with both \texttt{ionFR} and \texttt{spinifex} 
for ionospheric corrections. We demonstrate these limits in Fig.~\ref{fig:gagg_main_main}.
Let us first present the analysis for the primary TPPA with the \texttt{ionFR}-corrected data. Since the earlier PPPA analysis~\cite{PPTA:2024mgh} also used \texttt{ionFR} for ionospheric corrections, we can compare their sensitivities, focusing on the aspects reflecting the differences in observations conducted and pulsar groups engaged.
As shown in this figure, the TPPA analyses achieve their best sensitivities at $m_a \sim 5 \times 10^{-23}$ eV, where the ALDM signals oscillate in the PA residual time series with periods of several years, comparable to the observational timespan of the TPPA pulsars. This marks the turning point for TPPA detection: below this point the TPPA gradually loses its constraining power as the signal oscillation periods exceed the observational timespan, while above it, the signal strength weakens due to an approximately linear suppression of the large $m_a$ on the ALDM field amplitude.
Comparing to the PPPA limits, the TPPA sensitivities are visibly weaker below this point since its observational timespans are much shorter than those of the PPPA, namely $\lesssim~6$ years (TPPA) versus $\lesssim~18$ years (PPPA). However, their exclusion curves become intertwined above this mass point, yielding TPPA's surpassing in limits for mass intervals such as $ 3\times 10^{-22}<m_a < 10^{-21}\,$eV. 
This demonstrates the scientific potential of regular-pulsar PPAs enabled by new-generation observatories. Selecting tens of pulsars from a thousand-scale catalogue yields a sensitivity in the large $m_a$ regime comparable to an MSP-based PPA, with only one‑third of its observing time.

The TPPA analysis with \texttt{spinifex} exhibits a similar trend of sensitivity variation with respect to $m_a$, yielding an overall improvement in sensitivities compared to the \texttt{ionFR} case. These limits surpass the PPPA ones for a broad ALDM mass range, establishing new world-leading constraints on the ALDM Chern-Simons coupling within $10^{-23} {\rm eV}<m_a < 10^{-21}\,$eV. 
This outcome strongly aligns with the improved performance of the noise model for individual pulsars, achieved by applying the \texttt{spinifex} correction to the data~\cite{Sarkis:2025}, and indicates that developing a more accurate noise model - capable of better accounting for the complex ionospheric effects as well as general astronomical FR effects - represents an important direction for future work.



\begin{figure}
    \centering
    \includegraphics[width=\columnwidth]{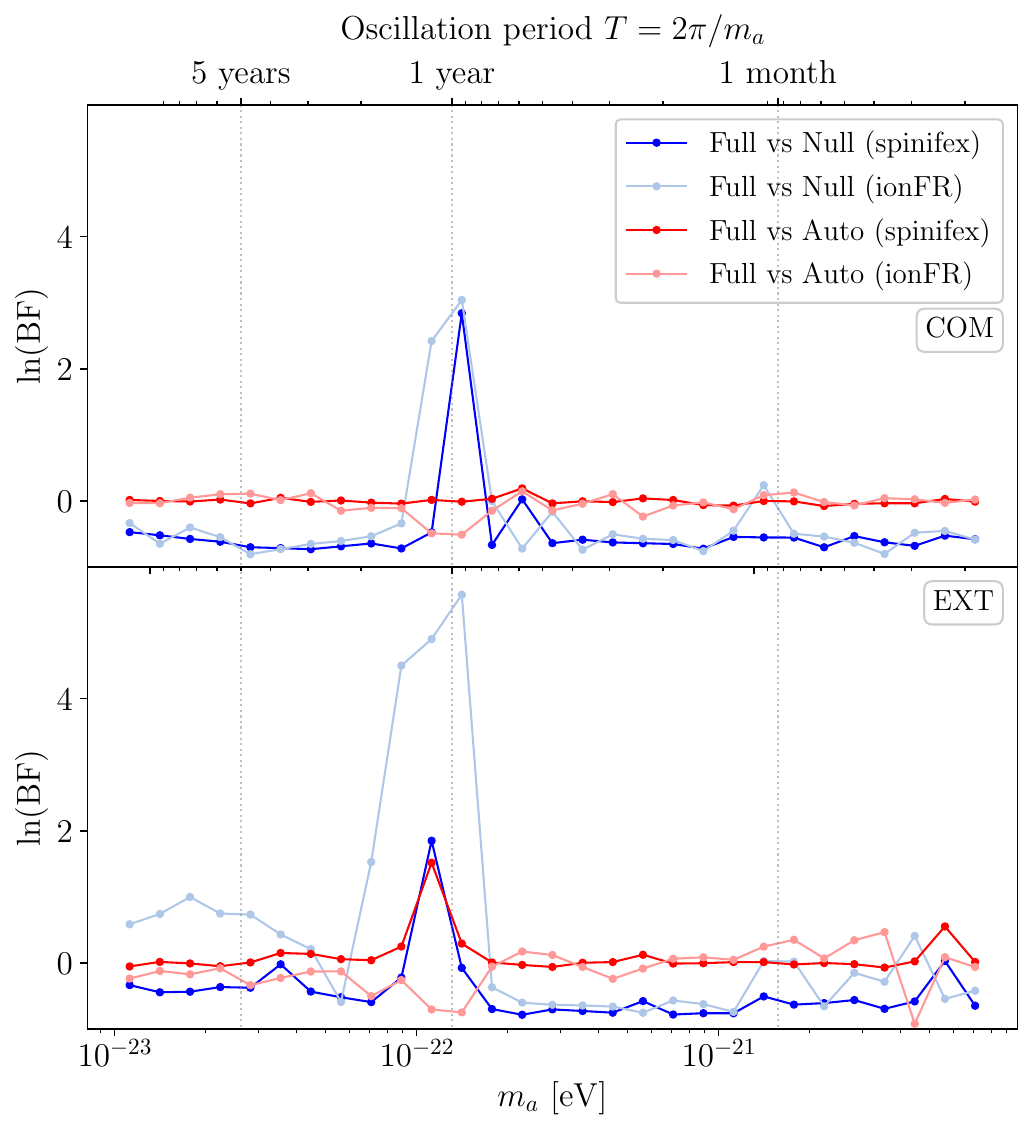}
    \caption{Bayes Factors indicating model significance for COM TPPA of pulsars (upper panel) and EXT TPPA (lower panel). 
    The ``Full vs Null'' curves correspond to $\mathrm{BF}_\mathrm{Null}^\mathrm{Full}$, and likewise for ``Full vs Auto''. 
    }
    \label{fig:lnBF_double_main}
\end{figure}

Despite these inspiring results in constraining the ALDM Chern-Simons coupling, anomalous gentle peaks appear on the one-year time scale for the TPPA exclusion curves. To recognize their nature, we compute two  Bayes factors, namely $\mathrm{BF}_\mathrm{Null}^\mathrm{Full}$ and $\mathrm{BF}_\mathrm{auto}^\mathrm{Full}$, for each analysis and demonstrate them as a function of $m_a$ in Fig.~\ref{fig:lnBF_double_main}. Here $\mathrm{BF}_\mathrm{Null}^\mathrm{Full}$ assesses the data preference of full-correlation signal model against the null-signal model, while $\mathrm{BF}_\mathrm{auto}^\mathrm{Full}$ examines the relevance of the ALDM-induced pulsar cross correlations to the observed anomalies in the data which are fully accounted for and turned off in the full- and auto-correlation signal models, respectively.     
These two Bayes factors can serve as key signatures of the ultralight ALDM signals~\cite{PPTA:2024mgh}. We consider the COM and EXT TPPAs instead, which are parts of the primary TPPA, for dissecting the roles played by different groups of pulsars.  

Figure~\ref{fig:lnBF_double_main} shows that the $\ln \mathrm{BF}_\mathrm{Null}^\mathrm{Full}$ baselines are below zero across most of the ALDM mass range, indicating a general preference of the data for the null-signal model. The analyses using \texttt{ionFR} produce peaks near $T=1\,$yr. However, their $\ln \mathrm{BF}_\mathrm{Auto}^\mathrm{Full}$ curves fluctuate downwards at this position for both COM and EXT TPPAs, implying no evidence for the ALDM signals. In the analyses with \texttt{spinifex}, the $\ln \mathrm{BF}_\mathrm{Null}^\mathrm{Full}$ curve also peaks at $T=1\,$yr, albeit less prominently. But at the same position, the $\ln \mathrm{BF}_\mathrm{Auto}^\mathrm{Full}$ curve is flat for the COM TPPA, while exhibiting minimal evidence only for the EXT TPPA. 
Thus, the observed anomalous peaks in Fig.\ref{fig:gagg_main_main} are more likely due to imperfect noise modelling.

\textit{\textbf{Conclusions.}---} 
In this \textit{Letter}, we present the first PPA of regular pulsars using data from MeerKAT's TPA Programme, to search for ultralight ALDM. The abundance of regular pulsars in astronomy and the precision of their polarization measurements, both competitive with MSPs, may significantly expand future research opportunities. Despite the relatively short observational period of the TPA data, we have established  constraints on the ALDM Chern-Simons coupling surpassing the current best limits for the range of $ 10^{-23}{\rm eV}<m_a<10^{-21}{\rm eV}$ and set a new baseline for future non-gravitational explorations of the ultralight ALDM.

The outcome remains limited due to the unsystematic removal of astronomical FR effects, despite the ionospheric subtraction with \texttt{spinifex}\cite{mevius_2025_15000430} and \texttt{ionFR}\cite{Sotomayor-Beltran:2013vma}. These effects can contribute to PA residuals through channels such as temporal variations of the IISM properties, the line of sight to pulsars, and local magnetic field and plasma profiles due to solar activities, in addition to ionospheric impacts. This challenge could be addressed by conducting multi-frequency-band analyses, as FR is frequency-sensitive while ALDM-induced CB is frequency-independent, and by correlating polarization data of regular pulsars and timing data of MSPs, as suggested in~\cite{Li:2025xlr}. With the upcoming large volume of polarization data from regular pulsars monitored over a decade with revolutionary precision by new-generation radio telescopes, we anticipate that the PPAs of regular pulsars will unlock even greater scientific potential in this area and beyond (see, e.g.,~\cite{Liang:2025vji}).

\textit{\textbf{Acknowledgments.}---}
We thank Xiaowei Ou for providing posterior chains for the dark matter halo modelling, and Roland Crocker, Geoff Beck, Chris Gordon and Steven Murray for helpful comments and suggestions on this work. Z.-Y. Yuwen acknowledges the support from the Program of China Scholarship Council Grant No. 202404910329, and is also supported by the Young Scientist Training (YST) Program at the APCTP through the Science and Technology Promotion Fund and Lottery Fund of the Korean Government. Y.-Z. Ma acknowledges the support from South Africa's National Research Foundation under Grants No.~150580, No.~CHN22111069370 and No.~ERC250324306141. 
J.R. is supported in part by the National Natural Science Foundation of China under Grant No. 12435005.
X.X. is funded by the grant CNS2023-143767. 
Grant CNS2023-143767 funded by MICIU/AEI/10.13039/501100011033 and by European Union NextGenerationEU/PRTR. We acknowledge computational cluster resources at the Centre for High-Performance Computing, Cape Town, South Africa.
The MeerKAT telescope is operated by the South African Radio Astronomy Observatory, which is a facility of the National Research Foundation, an agency of the Department of Science and Innovation. MeerTime data are housed and processed on the OzSTAR supercomputer at Swinburne University of Technology. Pulsar research at Jodrell Bank Centre for Astrophysics and Jodrell Bank Observatory is supported by a consolidated grant from the UK Science and Technology Facilities Council (STFC).

\section*{Supplementary Materials}
\setcounter{section}{0}
\setcounter{equation}{0}
\setcounter{figure}{0}
\setcounter{table}{0}
\renewcommand{\theequation}{S\arabic{equation}}
\renewcommand{\thefigure}{S\arabic{figure}}
\renewcommand{\thetable}{S\arabic{table}}

In the Supplementary Materials, we provide a detailed description of data preparation in Sec.~\ref{sec: data prepare}, the analysis methodology in Sec.~\ref{sec: methodology}, and additional analysis results in Sec.~\ref{sec: add res}. 

\section{Data preparation}\label{sec: data prepare}



\subsection{TPPA construction}


The TPA program has monitored 1,237 pulsars, many since February 2019. Their polarization time series are cleaned by removing outliers in $\Delta \mathrm{PA}$, which may result from systematics like radio interference (see~\cite{Sarkis:2025} for details). After excluding pulsars with fewer than twenty valid data points, 513 pulsars remain. From this collection, we construct one primary TPPA, using $50$ pulsars with the highest signal-to-noise  ratio (SNR) values in light intensity (termed the ``SNR'' group), and two auxiliary TPPAs, using 10 pulsars of SNR for each which have relatively compact and broad spatial distributions respectively (termed the ``compact (COM)'' and the ``extended (EXT)'' groups), for this study. Such an operation also excludes pulsars with poor noise modelling or anomalous Markov-Chain Monte Carlo (MCMC) results in noise analysis. More information on the polarisation time series of the TPA pulsars and their noise modelling can be found in~\cite{Sarkis:2025}.

\begin{figure} 
    \centering
    \includegraphics[width=0.98\linewidth]{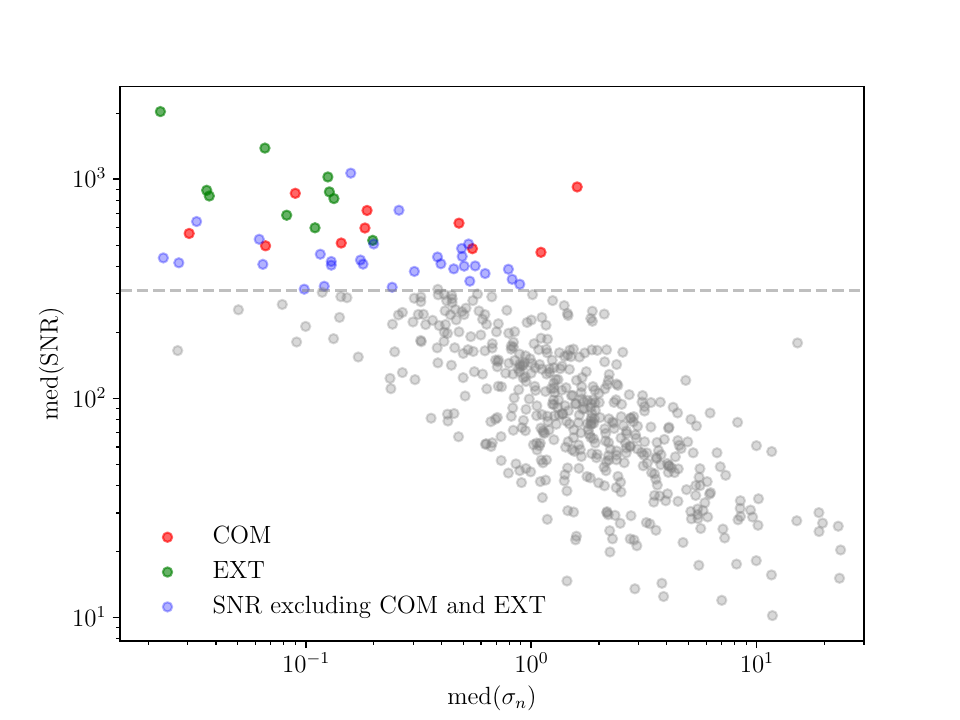}
    \caption{SNR median versus $\sigma_n$ median for $513$ cleaned TPA pulsars. The pulsars used for constructing the TPPAs are highlighted in red (COM), green (EXT) , and red+green+blue (SNR), respectively. The gray dashed line denotes the lower bound of SNR for these pulsars. }
    \label{fig: SNR_vs_sigma}
\end{figure}

The distribution of these pulsars on the med($\sigma_n$)-med(SNR) plane is demonstrated in Fig.~\ref{fig: SNR_vs_sigma}. Here med($\sigma_n$) and med(SNR) denote the medians of the PA observational error and the SNR respectively in the TPA pulsar time series. We summarize the basic information of the COM and EXT pulsars in Tab.~\ref{tab:groups}. Compared to the EXT pulsars, which have a mean distance of $4.98\,$kpc to the pulsars of the same group and the Earth, the COM pulsars, with a mean distance of $0.66\,$kpc, have a better chance to probe the signal spatial correlations of the ultralight ALDM in its median and high mass regimes.

\begin{table*}[ht]
    \begin{tabular}{c|c c c c c c c c c c c}
        \multirow{2}{*}{}
        Pulsar name & med(SNR) & $T_{\rm obs}$ & $N_{\mathrm{obs}}$  & $D_p$ & GC Dist& $\rho_p$ & med($\sigma_{n}^\mathrm{sp}$) & med($\sigma_{n}^\mathrm{io}$) & std($\delta \mathrm{PA^{sp}}$) & std($\delta \mathrm{PA^{io}}$) & RN\\ 
        & & [yr] &  & [kpc] & [kpc] & [$\mathrm{GeV\,cm^{-3}}$] & [deg] & [deg] & [deg] & [deg] &  \\
        \hline
        \textit{Compact TPPA} & & & \\  
        J1932+1059	& 922.0 & 5.51 & 75 & 0.31 & 8.29 & 0.4 &  1.63 & 1.66 & 2.48 & 3.35 & - \\
        J1752-2806  & 862.3 & 5.48 & 73 & 0.20 & 8.30 & 0.4 &  0.23 & 0.49 & 3.81 & 3.83 & - \\
        J1709-1640  & 720.1 & 6.20 & 79 & 0.56 & 7.96 & 0.5 &  0.30 & 0.50 & 6.13 & 4.24 & \checkmark \\
        J1534-5334  & 630.4 & 5.46 & 53 & 0.81 & 7.84 & 0.5 &  0.52 & 0.78 & 2.36 & 2.11 & - \\
        J1001-5507  & 598.9 & 5.45 & 49 & 0.30 & 8.45 & 0.4 &  0.25 & 0.60 & 3.03 & 3.20 & \checkmark \\
        J0630-2834  & 565.6 & 5.96 & 52 & 0.32 & 8.67 & 0.4 & 	0.22 & 0.44 & 1.46 & 1.37 & - \\
        J1705-1906  & 511.4 & 6.11 & 76 & 0.75 & 7.77 & 0.5 & 	0.26 & 0.49 &	1.38 & 2.31 & - \\
        J1456-6843  & 496.7 & 5.42 & 71	& 0.43 & 8.21 & 0.4 &  0.13 & 0.74 & 2.72 & 2.50 & -\\
        J1116-4122  & 482.0 & 5.51 & 73 & 0.28 & 8.44 & 0.4 &  0.59 & 0.76 &	2.36 & 2.40 & \checkmark \\
        J0953+0755  & 464.0 & 5.50 & 72 & 0.26 & 8.63 & 0.4 &  1.14 & 1.15 &	1.12 & 2.32 & - \\[1em]
        \textit{Extended TPPA}  & & & \\
        J1935+1616  & 2034.3 & 5.52 & 54 & 3.7 & 6.90 & 0.6 & 0.33 & 0.20 & 0.65 & 2.82 & - \\
        J1645-0317  & 1386.3 & 5.46 & 73 & 4.0 & 5.39 & 0.8 & 0.25 & 0.40 & 1.43 & 1.75 & - \\
        J1600-5044  & 1023.8 & 5.46 & 75 & 6.9 & 4.20 & 1.1 & 0.22 & 0.56 &	1.84 & 2.09 & - \\
        J1709-4429  & 889.9 & 6.11 & 160 & 2.6 & 6.06 & 0.7 & 0.18 & 0.59 & 1.29 & 1.09 & - \\
        J1820-0427  & 875.3	& 5.46 & 76 & 2.7 & 6.18 & 0.7 & 0.26 & 0.43 & 1.27 & 1.57 & -\\
        J1359-6038  & 838.3 & 6.00 & 76 & 5.0 & 6.42 & 0.7 & 0.14 & 0.66 &	1.06 & 1.11 & \checkmark \\
        J0820-1350  & 815.8 & 5.94 & 48 & 1.9 & 9.67 & 0.3 & 0.28 & 0.41 & 1.97 & 1.54 & - \\
        J1651-4246  & 685.1 & 6.11 & 81 & 5.2 & 3.87 & 1.2 & 0.19 & 0.52 &	1.16 & 1.51 & - \\
        J1829-1751  & 600.1 & 6.11 & 74 & 5.9 & 3.17 & 1.4 & 0.24 & 0.45 & 0.68 & 1.33 & \checkmark \\
        J1901+0331  & 526.0 & 5.50 & 56 & 7.0 & 5.15 & 0.9 & 0.33 & 0.35 & 0.82 & 2.33 & - \\
\hline
    \end{tabular}
    \caption{Information of the COM and EXT TPPA pulsars. Here ``med()'' and ``std()'' refer to the median and standard deviation of $\sigma_n$; 
    $D_p$ and ``GC Dist'' denote the pulsar distance to the Earth and Galactic Centre (GC) respectively; $\rho_p$ shows the ALDM density at the pulsar location (see Eq.~\eqref{eq: einasto}); $\delta \mathrm{PA^{sp}}$ and $\delta \mathrm{PA^{io}}$ represent PA residuals after the subtraction of ionospheric effects using the packages of  \texttt{spinifex} and \texttt{ionFR} respectively, and $\sigma_n^\mathrm{sp}$ and $\sigma_n^\mathrm{io}$ denote the uncertainties associated with these calculations in RM; and ``RN'' indicates the pulsars with a strong evidence of red noise in the Bayes noise analysis. See the text for details. }
    \label{tab:groups}
\end{table*}

\subsection{Pulsar position and distance}

Accurate information of pulsar position and distance is important for capturing the spatial correlations of ultralight ALDM signals in the analysis. The pulsar position is determined by its angular location on the celestial sphere $\hat{n}_p \equiv (\mathrm{RA}_p, \mathrm{DEC}_p)$ and distance to the Earth $D_p$ together. In this analysis, we will use the Australia Telescope National Facility (ATNF) Pulsar Catalog~\footnote{\url{https://www.atnf.csiro.au/research/pulsar/psrcat/} (Catalog Version 2.7.0)}~\cite{manchesterATNFPulsarCatalogue2005} to determine the position and distance of each pulsar. 

\begin{figure}
    \centering
    \includegraphics[width=\columnwidth]{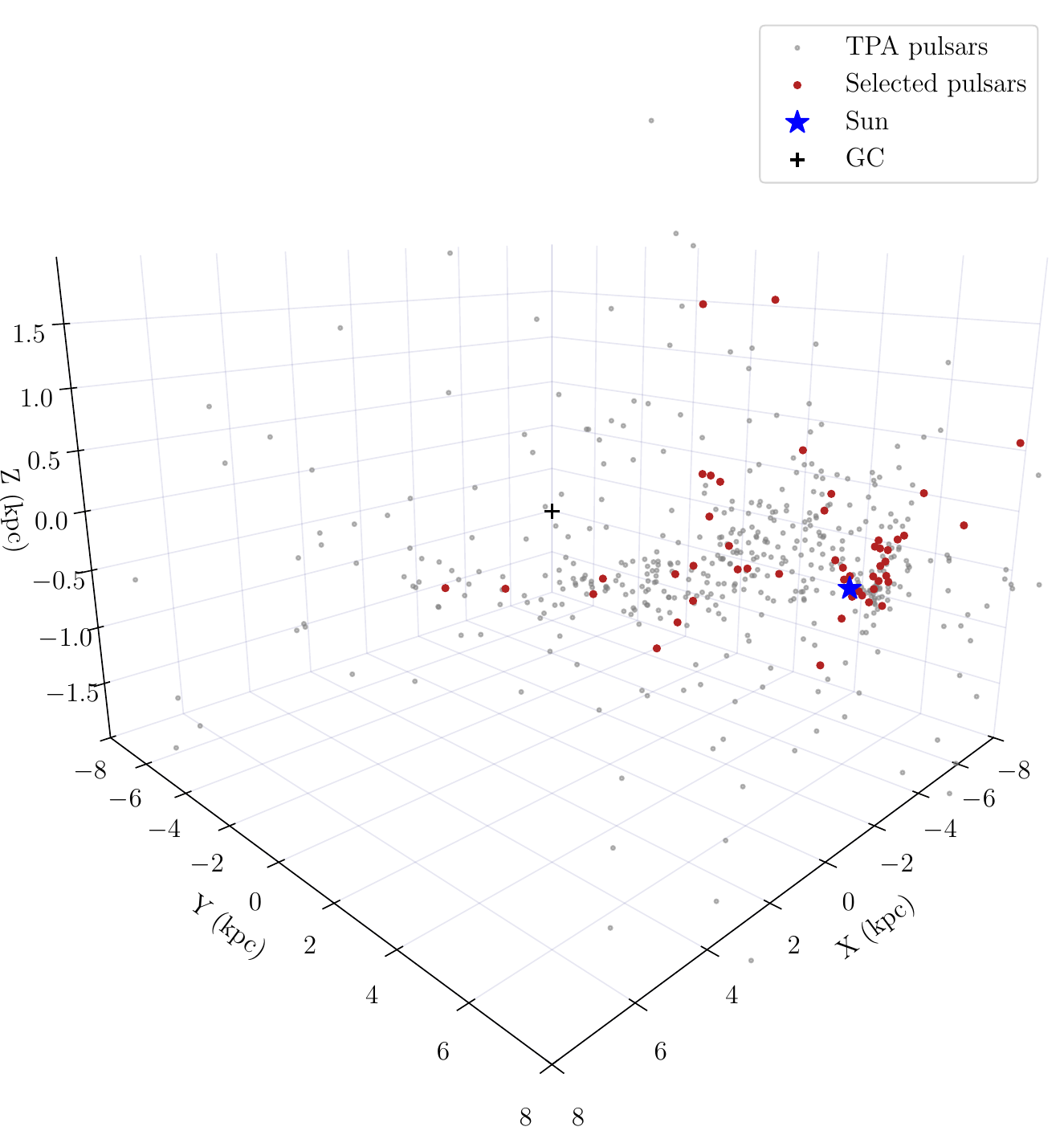}
    \caption{ Map of the TPA pulsars in Galactic Cartesian coordinates. Here, gray points represent the 513 pulsars selected from the TPA dataset, brown points indicate the 50 SNR pulsars, and the blue star represents the Sun.
    \label{fig:map}}
\end{figure}

As described in~\cite{manchesterATNFPulsarCatalogue2005}, the distance information of pulsars provided by this catalogue is derived quantities. So we determine their distance using a prioritization list of the available distance measures. We arrange the  priority to the values obtained from independent observations, such as those taken for the pulsars associated with supernova remnants or within globular clusters, which are generally thought to be a more reliable distance estimator. We then consider the values obtained using the information of annual parallax (PX), although we drop off the cases of large PX uncertainties where $\sigma_\mathrm{PX}>0.3\, \mathrm{PX}$, as in~\cite{NANOGrav:2023bts,PPTA:2024mgh}. If the two options are not available, we will take the values derived from Dispersion Measure ($\widetilde{\mathrm{DM}}$) variations using the YMW17 model~\cite{Yao:2017kcp} of the Galactic electron distribution. 
With this determination, a visualisation of the spatial distribution of TPA pulsars is given in Fig.~\ref{fig:map}.

For the convenience, we define a dimensionless variable $D_*\equiv D_p/D_0$, where $D_0$ is the central value of the measured distance. 
For the PX method, the pulsar distance and its $D_*$ error are given by
\begin{align}
    \frac{D_0}{\mathrm{kpc}} = \frac{\mathrm{mas}}{\mathrm{PX}}~,\quad \sigma_{D_*} = \frac{\sigma_{\mathrm{PX}}}{\mathrm{PX}}~,
\end{align}
with the $D_*$ prior modelled as
\begin{align}\label{eq: Dp prior}
    \ln \mathrm{Pr}(D_*) = -\frac{(D_*^{-1}-1)^2}{2\sigma_{D_*}^2} - 2\ln D_* + \mathrm{const.}~,
\end{align}
where the constant term is determined by the normalization condition $\int \mathrm{Pr}(x)\md x=1$. 


For the other two methods, the $D_*$ prior is modelled as a broken-half-Gaussian distribution, constructed from two half-Gaussian functions with a uniform distribution in the middle, namely
\begin{eqnarray}
    && \ln \mathrm{Pr}(D_*) = \mathrm{const.}  \nonumber \\
    && \quad  - \left\{
    \begin{array}{ll}
        (D_*-D_-)^2/(2\sigma^{2}_{\ast}) & ,~D_*<D_- \\
        0 & ,~D_-<D_*<D_+ \\
        (D_*-D_+)^2/(2\sigma^{2}_{\ast}) & ,~D_*>D_+
    \end{array}
    \right.
\end{eqnarray}
where $\sigma_*=0.2$ is the Gaussian distribution variation, and $D_\pm$ are the upper and lower bounds for the uniform distribution in the middle, corresponding to the relevant observational uncertainties. As above, the constant term is determined by the normalization condition.

Notably, the halo density profile for the ultralight ALDM is quite complex, since its wave interference may yield order unity density fluctuations on de Broglie scale~\cite{huiWaveDarkMatter2021}. Following common practices, we consider a simplified setup, using the spherically symmetric Einasto profile~\cite{ouDarkMatterProfile2024} as an approximation
\begin{equation}\label{eq: einasto}
    \rho_{\mathrm{Ein}}(r) = \dfrac{M_0}{4\pi r_{\rm s}^3}\exp\left[-\left(\frac{r}{r_{\rm s}}\right)^\alpha\right] \,,
\end{equation}
where $M_0$ is the normalisation mass for the halo, $r_{\rm s}$ is a reference radius, and $\alpha$ is a slope parameter. While the commonly-used NFW profile could be fitted for this study also, the data appears to favor the more cored Einasto profile with parameters $M_0 = 0.62^{+0.12}_{-0.11} \times 10^{11} \,\mathrm{M_{\odot}}$, $r_{\rm s} = 3.86^{+0.35}_{-0.38}\, \mathrm{kpc}$ and $\alpha = 0.91^{+0.04}_{-0.05}$~\cite{ouDarkMatterProfile2024}, which we have implemented in this work.

\section{Methodology}\label{sec: methodology}


Our TPPA analysis for detecting ultralight ALDM primarily adheres to standard PTA Bayesian analysis practices~\cite{Lentati:2013rla} and the recently developed PPPA analysis framework. An in-depth investigation on noise properties of the TPPA pulsars and their temporal trends in polarization is presented in~\cite{Sarkis:2025}. 


\subsection{Signal and noise modelling}\label{subsec: signal modelling}


The observed PA residuals of TPPA pulsars are modelled as a sum of  individual components~\cite{PPTA:2024mgh}:
\begin{align}
    \Delta\mathrm{PA}^\mathrm{obs} = \Delta\mathrm{PA}^\mathrm{ion} + \Delta\mathrm{PA}^\mathrm{IISM} + \Delta\mathrm{PA}^\mathrm{n} + \Delta\mathrm{PA}^a \, ,
\end{align}
where $\Delta\mathrm{PA}^a$ is the ALDM-induced PA residual, $\Delta\mathrm{PA}^\mathrm{ion}$ denotes  
the Earth's ionospheric contribution, thereby $\delta {\rm PA}\equiv \Delta\mathrm{PA}^\mathrm{obs}-\Delta\mathrm{PA}^\mathrm{ion}$ defining the {\it ionosphere-corrected data}, $\Delta\mathrm{PA}^\mathrm{IISM}$ encodes the effects of long-term variations along the line of sight to the pulsar in the Ionised InterStellar Medium (IISM), and $\Delta\mathrm{PA}^\mathrm{n}\equiv \Delta\mathrm{PA}^\mathrm{w} + \Delta\mathrm{PA}^\mathrm{r}$ represents impacts from random white ($\Delta\mathrm{PA}^\mathrm{w}$) and red ($\Delta\mathrm{PA}^\mathrm{r}$) noise. 




The signal for ultralight ALDM is a consequence of cosmological birefringence, with its magnitude determined by the field difference between the spacetime points of pulsar light emission and reception~\cite{Liu:2019brz,Liu:2021zlt}: 
\begin{align}\label{eq: signal}
    \Delta\mathrm{PA}^a \simeq g_{a\gamma\gamma}\left(a(\mathbf{x}_\mathrm{p},t_p) - a(\mathbf{x}_\mathrm{e},t_e)\right)~.
\end{align}
Here $a(\mathbf{x},t)$ represents a stochastic superposition of a vast number of particle waves on galactic scales~\cite{khmelnitskyPulsarTimingSignal2014,huiWaveDarkMatter2021}. In Bayes analysis, this signal can be characterized by the covariance matrix $C^a$ or the two-point correlation functions between the $\Delta\mathrm{PA}^a$ time series of pulsars $p$ and $q$~\cite{Liu:2021zlt}: 
\begin{align}
\begin{aligned}\label{eq: signal_covariance}
    C^a_{p,n;q,m} &= \langle \Delta\mathrm{PA}^a_{p,n}\, \Delta\mathrm{PA}^a_{q,m} \rangle \\  &= \frac{g_{a\gamma\gamma}^2 \rho_\mathrm{e}}{m_a^2} \bigg\{ \cos\left[ m_a(t_{p,n}-t_{q,m}) \right] \\
    & \quad~ + \sqrt{\frac{\rho_p \rho_q}{\rho_\mathrm{e}^2}} \cos\left[ m_a(t'_{p,n}-t'_{q,m}) \right] \frac{\sin y_{pq}}{y_{pq}} \\
    & \quad~ - \sqrt{\frac{\rho_p }{\rho_\mathrm{e}}} \cos\left[ m_a(t'_{p,n}-t_{q,m}) \right] \frac{\sin y_{\mathrm{e}p}}{y_{\mathrm{e}p}} \\
    & \quad~ - \sqrt{\frac{\rho_q}{\rho_\mathrm{e}}} \cos\left[ m_a(t_{p,n}-t'_{q,m}) \right] \frac{\sin y_{\mathrm{e}q}}{y_{\mathrm{e}q}} \bigg\} \\
    &\equiv S_a^2 \hat{C}^a_{p,n;q,m}~,
\end{aligned}
\end{align}
where $S_a \equiv g_{a\gamma\gamma}\sqrt{\rho_\mathrm{e}}/m_a$
is the signal characteristic strenth, $t_{p,n}$ and $t'_{p,n} = t_{p,n}- D_p$ are the reception and emission time of the $n$-th observation epoch for the $p$-th pulsar, and $D_p$ is the distance to the $p$-th pulsar. $y_{ij}=|\mathbf{x}_i-\mathbf{x}_j|/l_{\rm c}$ is dimensionless, comparing spatial separations of pulsars or between a pulsar and the Earth, and the ALDM coherence length $l_{\rm c}=1/(m_av_0)$ for a given mass $m_a$ and  velocity $v_0 \sim 10^{-3}$. Here $l_c$ is different from the ALDM de Broglie wavelength by a factor of $2\pi$. 
The temporal and spatial correlations of the ALDM signal, as manifestations of its wave nature, are encoded in the sinc and cosine factors, respectively. Notably, even if the signal's spatial correlations are suppressed for a large $m_a$ where $y_{ij} \gg 1$, the pulsar cross-correlation can still retain its value by receiving contributions from the signal's temporal correlations.



The random noise in the dataset includes white noise  accounting for the temporally uncorrelated pulse-to-pulse errors, and red noise accounting for the temporally correlated RM variances. 
As in~\cite{PPTA:2024mgh,NANOGrav:2023ctt}, we assume no random noise correlated between different pulsars. The covariance matrix is then composed of diagonal blocks:  
\begin{align}\label{eq: Cwhite}
    C^\mathrm{w}_{p,n;q,m} &= \left( \left(\mathrm{EFAC}_p \sigma_{p,n}^\mathrm{c}\right)^2 + \mathrm{EQUAD}_p^2 \right)\delta_{p,q}\delta_{m,n}~, \\
    \label{eq: Cred summation}
    C^\mathrm{r}_{p,n;q,m} &= \delta_{p,q} (S^\mathrm{r}_p)^2 \frac{\Delta f_p}{f_\mathrm{yr}} \sum_{k=1}^{k_{\mathrm{max},p}}  \left(\frac{k\Delta f_p}{f_\mathrm{yr}}\right)^{\Gamma_p} \nonumber \\ 
    & \quad\quad\quad\quad \times\cos\left[2\pi k\Delta f_p(t_{p,n} - t_{p.m})\right] ~,
\end{align}
where $\sigma_{p,n}^\mathrm{c} =  \sqrt{\sigma_{p,n}^2 + (\sigma_\mathrm{p,n}^{\rm io/sp})^2 \lambda^4}$ represents the observational error for the $p$-th pulsar at $n$-th observation epoch after ionospheric subtraction. 
Only a minority of TPPA pulsars show a strong evidence of red noise~\cite{Sarkis:2025}. For computational efficiency, we thus set $C^{\mathrm{r}}=0$ for pulsars with $\ln \mathrm{BF}^{\mathrm{r+w}}_{\mathrm{w}} < 2.3$. Here $\ln \mathrm{BF}^{\mathrm{r+w}}_{\mathrm{w}}$ compares the models with both white and red noises and with white noise only.  
Finally only six pulsars in the SNR TPPA are considered for red noise modelling, including three from the COM TPPA and two from the EXT TPPA (see Tab.~\ref{tab:groups}). 



Astronomical Faraday Rotation (FR), often characterised by the Rotation Measure (RM) of magnetised plasma in the form $\mathrm{FR} = \mathrm{RM}\,\lambda^2$, also contributes significantly to the noise in data due to its time variance. One source is the Earth's ionosphere,  
a layer of ionised molecules and electrons in the atmosphere through which the pulsar light traverses. 
Due to chaotic fluctuations in density caused by solar and atmospheric impacts and seasonal variations of sunlight luminosity across the Earth's atmosphere, precise modelling of ionospheric effects is difficult. The RM changes are usually calculated with computational tools~\cite{poraykoTestingAccuracyIonospheric2019,poraykoValidationGlobalIonospheric2023}. 
In this work we employ two packages, namely \texttt{spinifex}~\cite{mevius_2025_15000430} and \texttt{ionFR}~\cite{Sotomayor-Beltran:2013vma}, to calculate ionospheric FR, where the plasma information as the input is detailed in~\cite{Sarkis:2025}. The obtained RM results are then subtracted from the observed PA residuals.

Another important contribution to PA residuals arises from the IISM. Over timescales of years long, magnetic field and plasma density near the line-of-sight to each pulsar may vary, leading to a change in FR. To model this effect, we include a polynomial of time up to second order, as in~\cite{PPTA:2024mgh}: 
\begin{align}\label{eq: IISM}
\begin{aligned}
    \Delta\mathrm{PA}^\mathrm{IISM}_p &= \psi_{p}^{(0)} +  \psi_{p}^{(1)} \tilde t + \psi_{p}^{(2)} \tilde t^2 \, , 
\end{aligned}
\end{align}
where $\tilde t = t/{\rm 1\,yr}$.
$\Delta \mathrm{PA}^\mathrm{IISM}_p$ can be further decomposed as $M_p \psi_p$, with 
\begin{align}
    &M_p = \left(
    \begin{array}{ccc}
        1 & \tilde{t}_{p,1} & \tilde{t}_{p,1}^2 \\
        1 & \tilde{t}_{p,2} & \tilde{t}_{p,2}^2   \\
        \vdots & \vdots & \vdots \\
        1 & \tilde{t}_{p,N_p} & \tilde{t}_{p,N_p}^2 
    \end{array} \right)~, \label{eq:Mpsi1} \\
    &\psi_p^\transp = \left( \psi_p^{(0)},\psi_p^{(1)},\psi_p^{(2)} \right) \, .   \label{eq:Mpsi2}
\end{align}
%
%
Then the total $\Delta \mathrm{PA}^\mathrm{IISM}$ with $\mathcal{N}$ pulsars is given by
\begin{align}
    \Delta \mathrm{PA}^\mathrm{IISM} &= \left( \Delta \mathrm{PA}^\mathrm{IISM}_1, \Delta \mathrm{PA}^\mathrm{IISM}_2, \dots, \Delta \mathrm{PA}^\mathrm{IISM}_\mathcal{N} \right)^\transp = M\psi~, \\
    \psi^\transp &= \left( \psi^\transp_1, \psi^\transp_2, \dots, \psi^\transp_\mathcal{N} \right)~,\\
    M &= \mathrm{Bdiag}(M_1, M_2, \dots, M_\mathcal{N})~,
\end{align}
where $\mathrm{Bdiag}$ stands for block diagonal matrices.  

\subsection{Bayesian analysis}\label{subsec: Bayesian analysis}


\subsubsection{Priors and likelihoods}

The covariance matrix for the ALDM signal are characterized by the  parameters: the distance to each pulsar $D_p$, the characteristic signal magnitude $S_a$, and the ALDM particle mass $m_a$. The prior for $D_p$ has been given in Eq.~\eqref{eq: Dp prior}, while the prior for $S_a$ and $m_a$ are modelled as uniform distributions: 
\begin{align}
    \mathrm{Pr}(\log_{10}(S_a/\mathrm{rad})) &= U[-8,2]~, \\
    \mathrm{Pr}(\log_{10}(m_a/\mathrm{eV})) &= U[m_{a}^{\mathrm{min}},m_{a}^{\mathrm{max}}]~,
    \label{eq: ma prior}
\end{align}
where $m_{a}^{\mathrm{min}}$ and $m_{a}^{\mathrm{max}}$ are defined by the boundaries of the mass bin being analyzed. Note all mass bins in our analysis have a width of $\Delta \log_{10} (m_a/\mathrm{eV}) = 0.1$. 

For each pulsar $p$, there are four variables parametrizing its random noise: the white noise scaling factor $\mathrm{EFAC}_p$, the additional quadrature error $\mathrm{EQUAD}_p$, the red noise amplitude $S^{\mathrm{r}}_p$ and power-law index $\Gamma_p$. The priors of these parameters are given by
\begin{align}
\begin{aligned}
    \mathrm{Pr}(\log_{10}(\mathrm{EFAC}_p)) = U[-2,2]~\, , \\ 
    \mathrm{Pr}(\log_{10}(\mathrm{EQUAD}_p/\mathrm{rad})) = U[-8,2]~ \, , \\
    \mathrm{Pr}(\log_{10}(S^\mathrm{r}_p/\mathrm{rad})) = U[-8,2]~ \, , \\
    \mathrm{Pr}(\Gamma_p) = U[-8,0]~ \, .
\end{aligned}
\end{align}
Here the upper bound for $\Gamma_p$ is set to $0$ since the integral in the covariance matrix of red noise (see Eq.\eqref{eq: Cred summation}) converges only for a negative $\Gamma_p$. 
Additionally, there are three parameters characterizing the IISM effects: $\psi^{(i)}_p$ with $i=0,1,2$, 
which will be treated as nuisance parameters and integrated out from the likelihood.

The likelihood for the Bayesian analysis is given by
\begin{align}\label{eq: lnL}
    \ln \mathcal{L} =& -\frac{1}{2}\left(\delta \mathrm{PA} - M\psi\right)^{\transp} C^{-1} \left(\delta \mathrm{PA} - M\psi\right) - \frac{1}{2}|2\pi C| ,
\end{align}
where $\delta\mathrm{PA}$ represents the data after the ionospheric corrections. Different from~\cite{Sarkis:2025},  the covariance matrix here includes the $C^a$ terms in Eq.~(\ref{eq: signal_covariance}) in addition to $C^\mathrm{w}+C^\mathrm{r}$, to account for the ALDM-induced correlations in data.  
%
%
We then have the marginalized likelihood by integrating out the the three IISM parameters $\psi$ for each pulsar: 
\begin{align}\label{eq: lnLm}
\begin{aligned}
    \ln \mathcal{L}_\mathrm{m} &=  \frac{1}{2}(\delta\mathrm{PA}^c)^\transp C_M^{-1} \delta\mathrm{PA}^c  - \frac{1}{2}\ln |2\pi C_M|\\
    &\quad~ - \frac{1}{2}\delta\mathrm{PA}^\transp C^{-1} \delta\mathrm{PA} - \frac{1}{2}\ln |2\pi C| + \mathrm{const.},
\end{aligned}
\end{align}where $C_M=M^\transp C^{-1} M$ and $\delta\mathrm{PA}^c=M^\transp C^{-1}\delta\mathrm{PA}$.
Here the constant term contributes to normalization only and hence can be dropped. This likelihood thus relies on $5\mathcal{N} + 2$ parameters: ($\mathrm{EFAC}_p$, $\mathrm{EQUAD}_p$, $S^\mathrm{r}_p$, $\Gamma_p$, $D_p$) for each pulsar, with $p=1$, 2, $... ...$, $\mathcal{N}$, and ($S_a$, $m_a$) for the ALDM signal.   


Bayes Factors (BF) in the analysis are computed, using the product-space method~\cite{lodewyckxTutorialBayesFactor2011}. 
This transdimensional scheme encompasses the models to compare into one hybrid model, by introducing a hyper-parameter $n$ with a uniform prior $U[-1,1]$. The hybrid likelihood is then given by 
\begin{align}
    \mathcal{L}(n,\vartheta|d) = \left\{\begin{array}{lc}
        \mathcal{L}_{\mathrm{model1}}(\vartheta|d) ~,&-1\leq n <0~, \\
        \mathcal{L}_{\mathrm{model2}}(\vartheta|d) ~,&0\leq n < 1 ~, 
    \end{array}
    \right.
\end{align}
where $\vartheta$ is Cartesian product of the parameter sets of model 1 and model 2, and $d$ denotes pulsar data. The BF is estimated as 
\begin{align}
    \mathrm{BF}_{\mathrm{model1}}^{\mathrm{model2}} = \frac{N_{\mathrm{model2}}}{N_{\mathrm{model1}}}~\, ,
\end{align}
where $N_{\mathrm{model}}$ is the number of MCMC samples drawn for each model from the hybrid likelihood.


\subsubsection{Decomposition of covariance matrices}


To calculate the inverse of the covariant matrices $C$ and $C_M$, we decompose their individual components first and then apply the Woodbury Identity, as in~\cite{PPTA:2024mgh}. Below let us examine these components one by one. 

$C^\mathrm{w}$ (see Eq.~(\ref{eq: Cwhite})) is diagonal, and no further decomposition is needed. $C^\mathrm{r}$ (see Eq.~\eqref{eq: Cred summation}) is decomposed as
\begin{align}
    C^\mathrm{r}_{p,n;q,m} &= \delta_{p,q} (S^\mathrm{r}_p)^2  (F^\mathrm{r}_{p,n})^\transp \Phi^\mathrm{r}_{p} F^\mathrm{r}_{p,m}~,
\end{align}
where 
\begin{widetext}
\begin{align}
    \Phi^\mathrm{r}_p &= \frac{\Delta f_p}{f_\mathrm{yr}} \mathrm{Bdiag}\left( \left( \frac{\Delta f_p}{f_\mathrm{yr}} \right)^{\Gamma_p}\mathbf{I}_{2\times 2},\left( \frac{2\Delta f_p}{f_\mathrm{yr}} \right)^{\Gamma_p}\mathbf{I}_{2\times 2},\dots, \left( \frac{k_{\mathrm{max},p}\Delta f_p}{f_\mathrm{yr}} \right)^{\Gamma_p}\mathbf{I}_{2\times 2}\right)~,\\
    F^\mathrm{r}_{p,n} &= \mathrm{diag}\left(\cos\left(2\pi\frac{t_{p,n}}{T_p}\right),\sin\left(2\pi\frac{t_{p,n}}{T_p}\right), \dots,\cos\left(2\pi\frac{k_{\mathrm{max},p}t_{p,n}}{T_p}\right),\sin\left(2\pi\frac{k_{\mathrm{max},p}t_{p,n}}{T_p}\right) \right)^\transp~\, 
\end{align}
\end{widetext}
have a dimension of $2k_{\mathrm{max},p}\times 2k_{\mathrm{max},p}$ and $2k_{\mathrm{max},p}\times 1$, respectively. 
Then, defining a $2k_{\mathrm{max},p}\times N_p$ matrix $F^\mathrm{r}_{p}=(F^\mathrm{r}_{p,1},\dots,F^\mathrm{r}_{p,N_p})$, we have a $N_p\times N_p$ red noise covariance matrix
\begin{eqnarray}
C^\mathrm{r}_p = (S^\mathrm{r}_p)^2  (F^\mathrm{r}_{p})^\transp \Phi^\mathrm{r}_{p} F^\mathrm{r}_{p}  \label{eq:cm_rn}
\end{eqnarray}
for individual pulsars and a $N_\mathrm{tot}\times N_\mathrm{tot}$ block diagonal matrix
\begin{align}
    C^\mathrm{r} = \mathrm{Bdiag}(C^\mathrm{r}_1, C^\mathrm{r}_2, \dots, C^\mathrm{r}_\mathcal{N}) 
\end{align}
for $\mathcal{N}$ pulsars, with $N_\mathrm{tot}=\sum_{p=1}^\mathcal{N} N_p$.

The ALDM signal covariance matrix $C^a$~\cite{Liu:2021zlt} can be decomposed as 
\begin{align}
    C^a_{p,n;q,m} = S_a^2 (F^a_{p,n})^\transp \Phi^a_{p,q} F^a_{q,m}~,
\end{align}
where 
\begin{align}\label{eq: Phi pq}
    \Phi^a_{p,q}=\left(\begin{array}{cc}
        \Phi^{cc}_{p,q} & \Phi^{cs}_{p,q} \\
        \Phi^{sc}_{p,q} & \Phi^{ss}_{p,q}
    \end{array}\right)~,
    \quad
    F^a_{p,n}=\left(\begin{array}{c}
        \cos(m_a t_{p,n})  \\
        \sin(m_a t_{p,n})
    \end{array}\right)~,
\end{align}
with 
\begin{align}
    &\Phi^{cc}_{p,q}=\Phi^{ss}_{p,q} = 1 + \sqrt{\tilde{\rho}_p \tilde{\rho}_q} \cos\left[m_a(D_p-D_q)\right] \\
    &\quad -\sqrt{\tilde{\rho}_p}\mathrm{sinc}(y_{\mathrm{e}p})\cos(m_a D_p) -\sqrt{\tilde{\rho}_q}\mathrm{sinc}(y_{\mathrm{e}q})\cos(m_a D_q)~, \nonumber\\
    &\Phi^{sc}_{p,q}=-\Phi^{cs}_{p,q} =  \sqrt{\tilde{\rho}_p \tilde{\rho}_q} \sin\left[m_a(D_p-D_q)\right] \\
    &\quad -\sqrt{\tilde{\rho}_p}\mathrm{sinc}(y_{\mathrm{e}p})\sin(m_a D_p) -\sqrt{\tilde{\rho}_q}\mathrm{sinc}(y_{\mathrm{e}q})\sin(m_a D_q)~, \nonumber
\end{align}
and $\tilde{\rho}_p =\rho_p/\rho_e$. 
Again defining a $2\times N_p$ matrix $F^a_p=(F^a_{p,1},\dots,F^a_{p,n})$, we can write $C^a$ in a more compact form
\begin{eqnarray}
    C^a=(S_a)^2 (F^a)^\transp \Phi^a F^a \, ,  \label{eq:cm_sig}
\end{eqnarray}
where $\Phi^a$ is a $2\mathcal{N}\times 2\mathcal{N}$ block matrix composed of $\mathcal{N}^2$ $2\times 2$ submatrices given in  Eq.~\eqref{eq: Phi pq}:
\begin{align}
    \Phi^a = \left(\begin{array}{cccc}
        \Phi^a_{1,1} & \Phi^a_{1,2} & \dots & \Phi^a_{1,\mathcal{N}} \\
        \Phi^a_{2,1} & \Phi^a_{2,2} & \dots & \Phi^a_{2,\mathcal{N}} \\
        \vdots & \vdots & \ddots & \vdots\\
        \Phi^a_{\mathcal{N},1} & \Phi^a_{\mathcal{N},2} & \dots & \Phi^a_{\mathcal{N},\mathcal{N}}
    \end{array}\right)~, 
\end{align}
and $F^a$ is given as $\mathrm{Bdiag}(F^a_1,\dots,F^a_\mathcal{N})$.

In view of the similar structure between their decompositions in Eq.~(\ref{eq:cm_sig}) and Eq.~(\ref{eq:cm_sig}), we can assemble the covariance matrices of red noise and ALDM signal as  $F^\transp \Phi F \equiv C^\mathrm{r}+C^a$, where $\Phi$ and $F$ are given by 
\begin{align}\label{eq: def Phi}
    \Phi &= \mathrm{Bdiag}\left((S_a)^2 \Phi^a, (S^\mathrm{r}_1)^2 \Phi^\mathrm{r}_1, (S^\mathrm{r}_2)^2 \Phi^\mathrm{r}_2,\dots, (S^\mathrm{r}_\mathcal{N})^2 \Phi^\mathrm{r}_\mathcal{N} \right)~,
\end{align}
\begin{align}\label{eq: def F}
    F= \left(\begin{array}{cccc}
        F^a_1 & & & \\
        & F^a_2 & & \\
        & & \ddots & \\
        & & & F^a_\mathcal{N} \\ 
        F^\mathrm{r}_1 & & & \\
        & F^\mathrm{r}_2 & & \\
        & & \ddots & \\
        & & & F^\mathrm{r}_\mathcal{N}
    \end{array}\right)~.
\end{align}
Then we have the total covariance matrix  
\begin{eqnarray}
    C=C^\mathrm{w} + F^\transp \Phi F \, .
\end{eqnarray}

The inverse of $C$ can be calculated using Woodbury Identity:
\begin{align} \label{eq: C inverse}
C^{-1} &= \left(C^\mathrm{w} + F^\transp \Phi F\right)^{-1} \\
    &= (C^\mathrm{w}_\mathrm{sqr})^{-1} (\mathbf{I} + \tilde{F}^\transp \Phi \tilde{F})^{-1} (C^\mathrm{w}_\mathrm{sqr})^{-1} \nonumber\\
    &= (C^\mathrm{w}_\mathrm{sqr})^{-1} \left(\mathbf{I} - \tilde{F}^\transp(\Phi^{-1}+\tilde{F}\tilde{F}^\transp)^{-1}\tilde{F} \right) (C^\mathrm{w}_\mathrm{sqr})^{-1}~,  \nonumber
\end{align}
where $\tilde{F} = F(C^\mathrm{w}_\mathrm{sqr})^{-1}$ is the whitened $F$-matrix, and $C^\mathrm{w}_\mathrm{sqr}$ satisfies $C^\mathrm{w} = (C^\mathrm{w}_\mathrm{sqr})^2$ and can be calculated by taking the square root on each entries of the diagonal $C^{\rm w}$. 
%
%
With this decomposition, the calculation of $C^{-1}$ is converted to the calculation of $\Phi^{-1}$, which has a dimension     $N_\mathrm{tot}=\sum_p N_p$, much smaller than that of $C^{-1}$, namely $2(\mathcal{N} + \sum_p k_{\mathrm{max},p})$. 
Additionally, the whitening of $F$, which leverages the diagonal property of $C^\mathrm{w}$, simplifies the matrix multiplication in Eq.~\eqref{eq: C inverse}, and can speed up the calculations also. The determinant of $C$ finally can be evaluated as
\begin{align}
    |C|=|\Phi^{-1} + \tilde{F}\tilde{F}^\transp| \cdot|\Phi| \cdot |C^\mathrm{w}|~.
\end{align}
Similarly we can whiten $M$ by defining $\tilde{M} = (C^\mathrm{w}_\mathrm{sqr})^{-1} M$, and finally have
\begin{align}
    &~~C_M = \tilde{M}^\transp \left(\mathbf{I} - \tilde{F}^\transp(\Phi^{-1}+\tilde{F}\tilde{F}^\transp)^{-1}\tilde{F} \right) \tilde{M} ~,\\
    &\begin{aligned}
        |C_M| &= |MM^\transp|\cdot|C|^{-1} \\ 
        &= |\tilde{M}\tilde{M}^\transp|\cdot|\Phi^{-1} + \tilde{F}\tilde{F}^\transp|^{-1} \cdot|\Phi|^{-1}~. 
    \end{aligned}
\end{align}

\section{More analysis results}\label{sec: add res}

\subsection{Limits from individual pulsars and pulsar arrays}

\begin{figure*}
    \centering
    \includegraphics[width=0.95\textwidth]{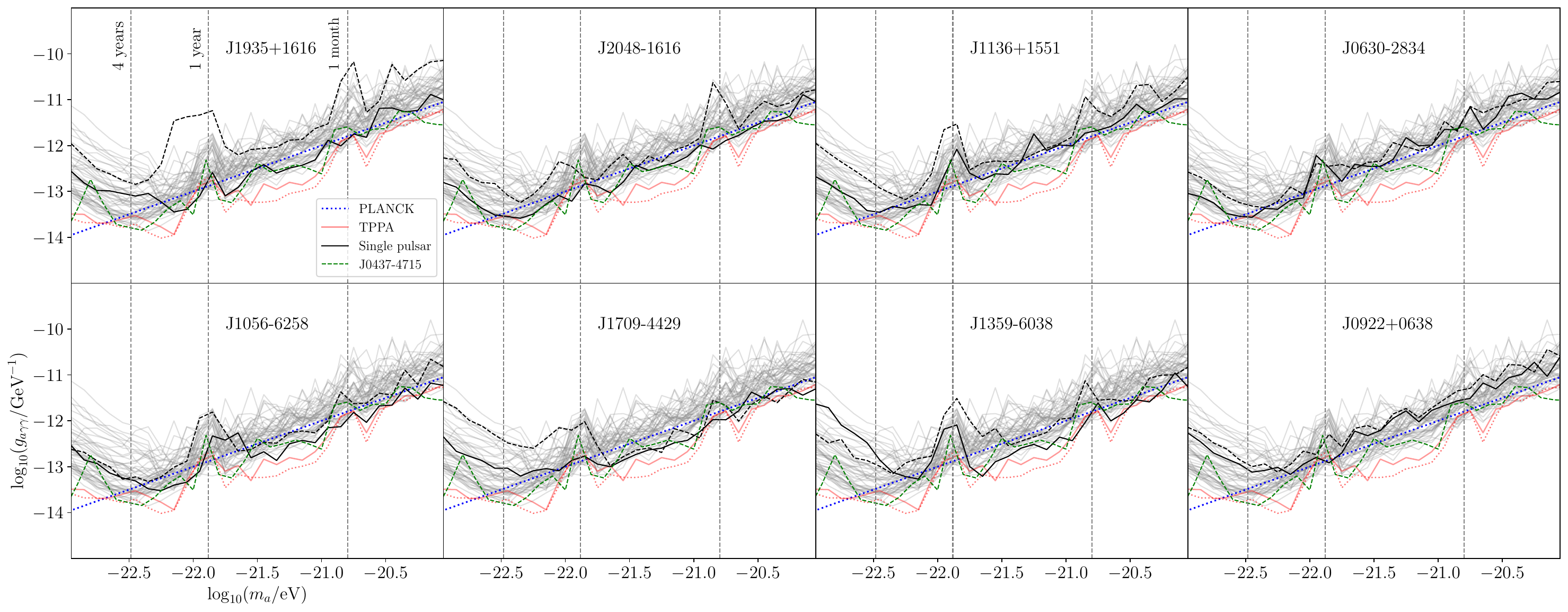}
    \caption{$95\%$ upper limits of eight individual SNR pulsars with the smallest med($\sigma_n$) on the ALDM Chern-Simons coupling $\gagg$ as a function of $m_a$.   
    The results obtained with \texttt{ionFR} and \texttt{spinifex} ionospheric subtractions are denoted as black-dashed and black-solid lines, while those set by the 
    remaining SNR pulsars, with the \texttt{spinifex} applied, are presented as grey lines. 
    For reference, we show the constraints from the  PPPA MSP J0437-4715~\cite{PPTA:2024mgh} as dashed green lines. We also present the major TPPA results, with \texttt{spinifex} applied, as red lines (the line styles of ``solid'' and ``dashed'' denote the signal models with full correlation and auto-correlation, respectively), and the CMB limits obtained using PLANCK data~\cite{fedderkeAxionDarkMatter2019} as blue-dashed lines. }
    \label{fig: SPA}
\end{figure*}


\begin{figure*}
    \centering
    \includegraphics[width=0.9\textwidth]{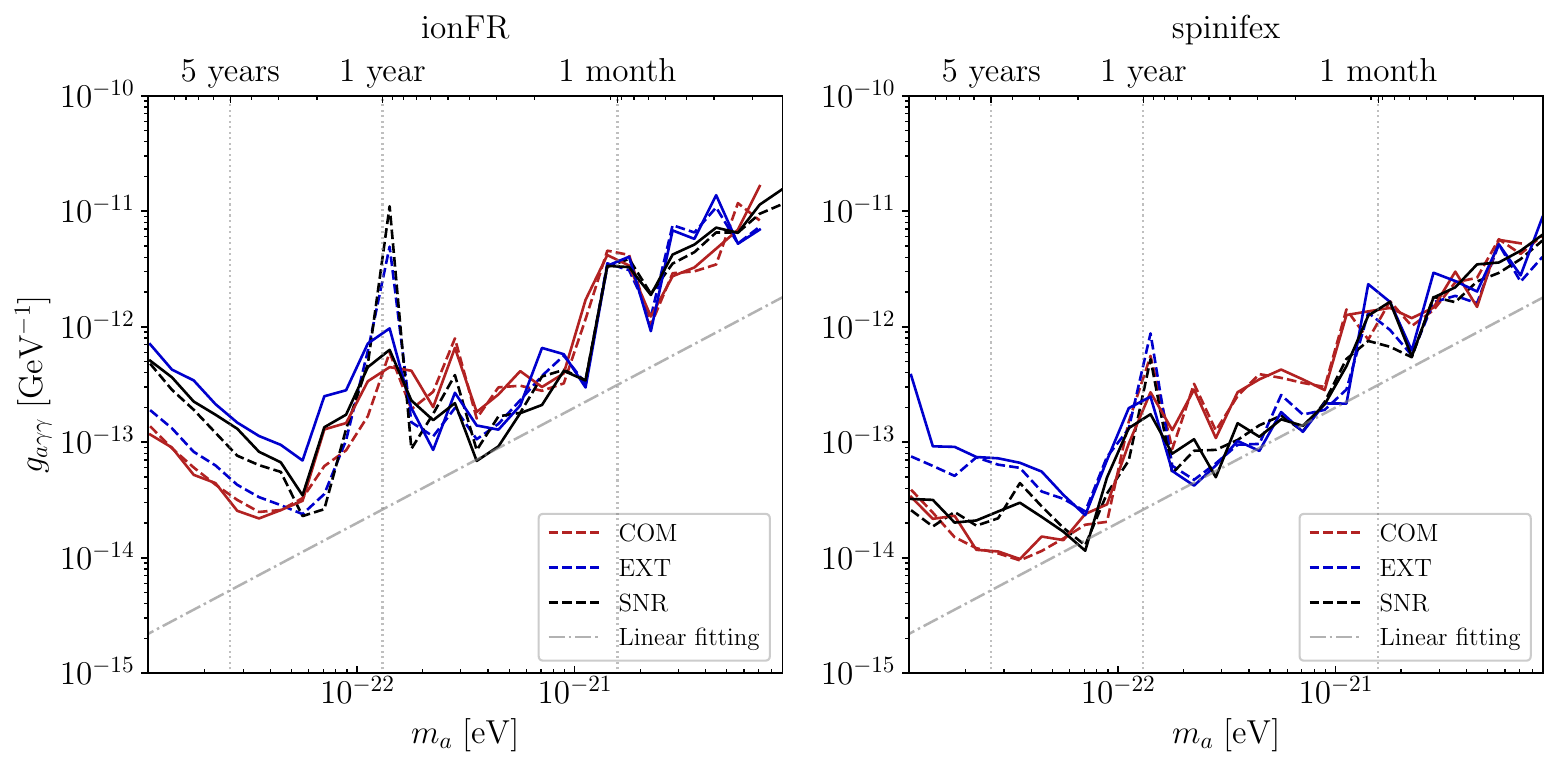} \\
    \includegraphics[width=0.9\textwidth]{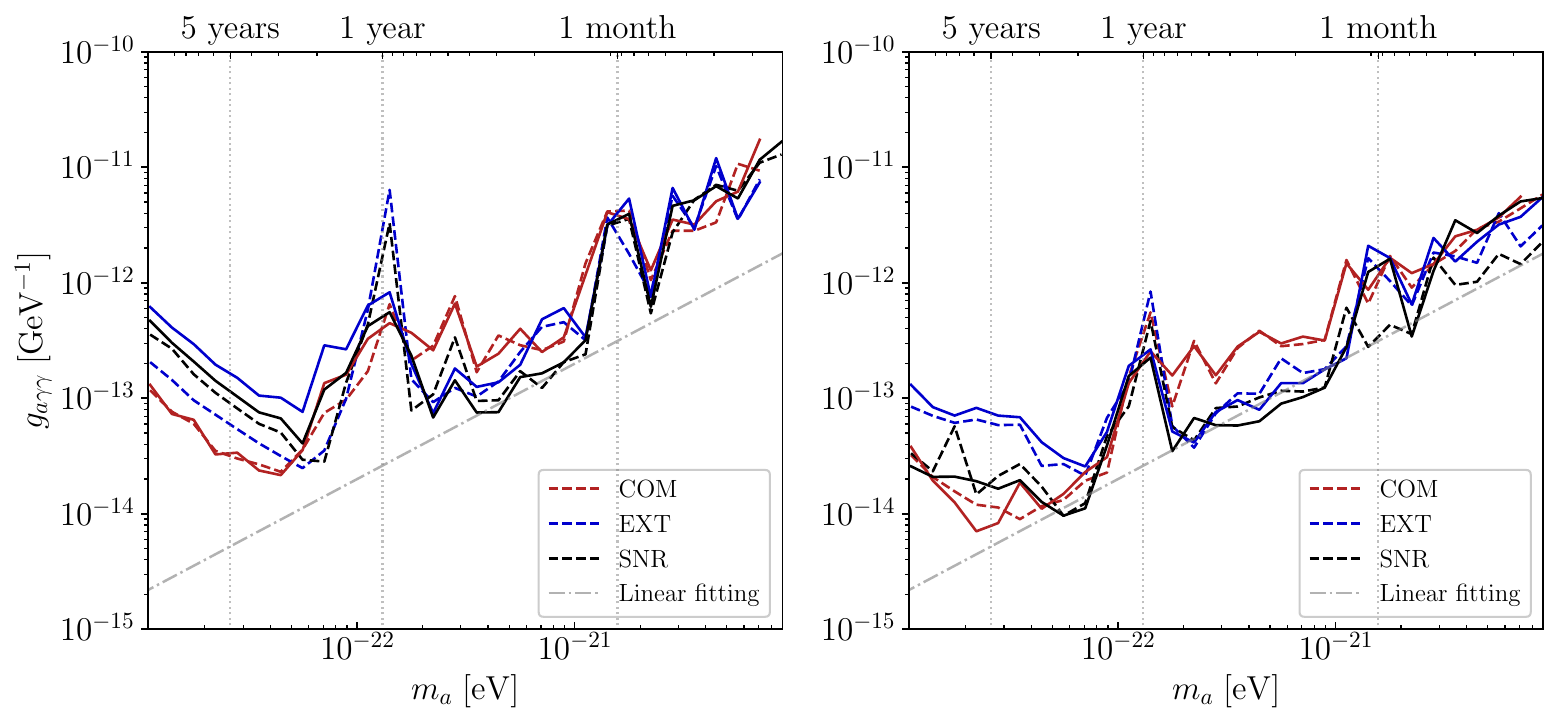}
    \caption{95\% upper limits of the three TPPAs on the ALDM Chern-Simons coupling $g_{a\gamma\gamma}$ as a function of $m_a$. These limits are obtained by employing the analysis strategies:  with \texttt{ionFR} (left column) and   \texttt{spinifex} (right column) ionospheric subtractions, with full-correlation (top row) and auto-correlation-only (bottom row) signal models, and without (solid lines) and with (dashed lines) the harmonic terms for noise modelling in Eq.~\eqref{eq: IISM}. The dot-dashed lines correspond to the linear fitting function $g_{a\gamma\gamma}\sim 10^{-13.3} \left(\frac{m_a}{10^{-22}\mathrm{eV}}\right)~\mathrm{GeV}^{-1}$ provided in the main text. }
    \label{fig:res of all groups}
\end{figure*}

We present in Fig.~\ref{fig: SPA} the $95\%$ upper limits of eight individual SNR pulsars with the smallest med($\sigma_n$) on the ALDM Chern-Simons coupling $\gagg$. For the analysis with the \texttt{ionFR} package applied, two pulsars, namely J1056-6258 and J1709-4429, exhibit a potential of  surpassing the CMB measurements in~\cite{fedderkeAxionDarkMatter2019} for certain mass ranges. This allows their performance to approach that of J0437-4715, the PPPA MSP with the smallest med($\sigma_n$), for $m_a \gtrsim 5\times 10^{-23}\,$eV, despite their much shorter observational timespan ($\sim 6\,$yrs vs. $\sim 18\,$yrs). The application of \texttt{spinifex} for ionospheric subtractions further improves the sensitivities of the eight pulsars universally. Several additional pulsars, such as J1935+1616, J2048+1616 and J1359-6038, also demonstrate a sensitivity comparable to that of the J0437-4715. This highlights the advantage of regular pulsars' large population in securing high-quality pulsars for constructing PPAs. Finally, with the collective power of pulsars, the primary TPPA enhances the exclusion limits to a level exceeding the PPPA ones for a broad ALDM mass range.

To recognize the contributions from different groups of pulsars, we demonstrate in Fig.~\ref{fig:res of all groups} the $95\%$ upper limits on the ALDM Chern-Simons coupling $\gagg$ achieved by the primary (SNR) and auxiliary (COM and EXT) TPPAs. The \texttt{spinifex}'s application yields an overall improvement to the limits set by the \texttt{ionFR}, as expected. In either case, the COM TPPA exhibits a better sensitivity than the EXT one for the mass regime of $m_a < 10^{-22}~\mathrm{eV}$. This could be attributed to the COM TPPA's stronger capability of detecting signal spatial correlations. 
Differently, the EXT TPPA yields slightly better constraints for the mass range of $\sim 10^{-22}-10^{-21}\,$eV, which could benefit from the relatively small observational errors of the EXT pulsars as well as the slightly larger local DM energy density around them. Then, by combining the COM and EXT TPPAs and the remaining 30 pulsars, the SNR TPPA generates the limits which, when the \texttt{spinifex} is applied, follow the sensitivity trends of the COM and EXT TPPAs for $m_a<10^{-22}\,$eV and $m_a>10^{-22}\,$eV, respectively.

\subsection{Additional harmonic noise components}

\begin{figure*}
    \centering
    \includegraphics[height=0.32\textwidth]{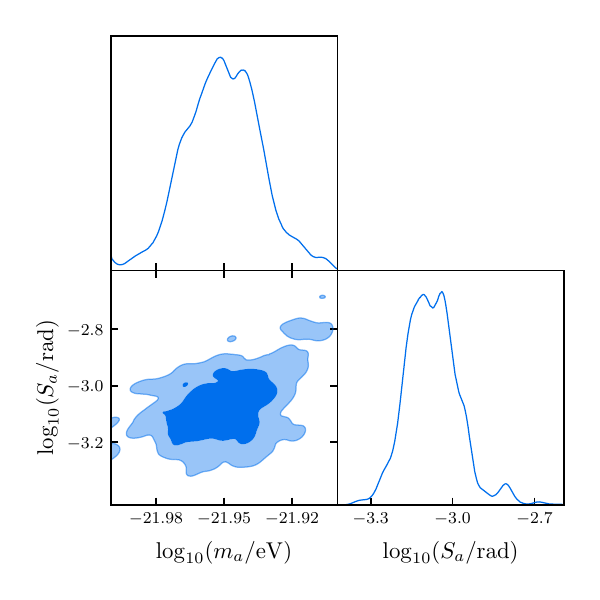}
    \includegraphics[height=0.32\textwidth]{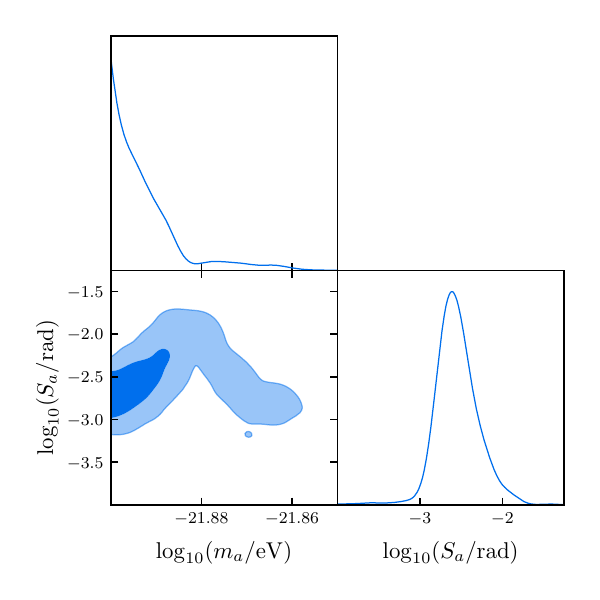}
    \includegraphics[height=0.32\textwidth]{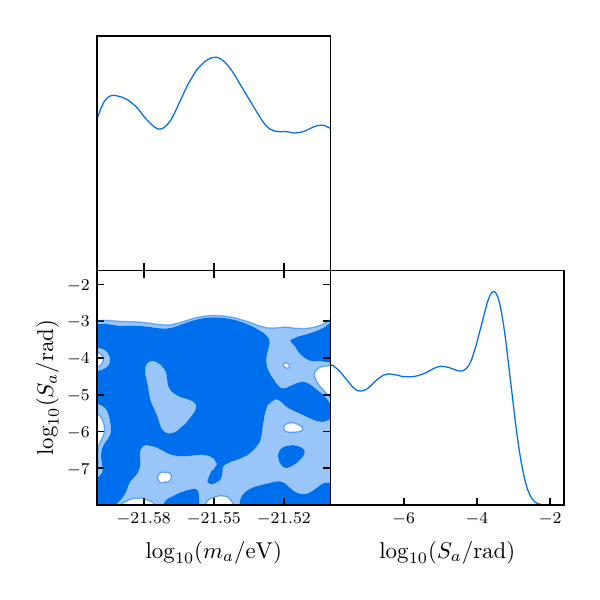} \\
    \includegraphics[height=0.32\textwidth]{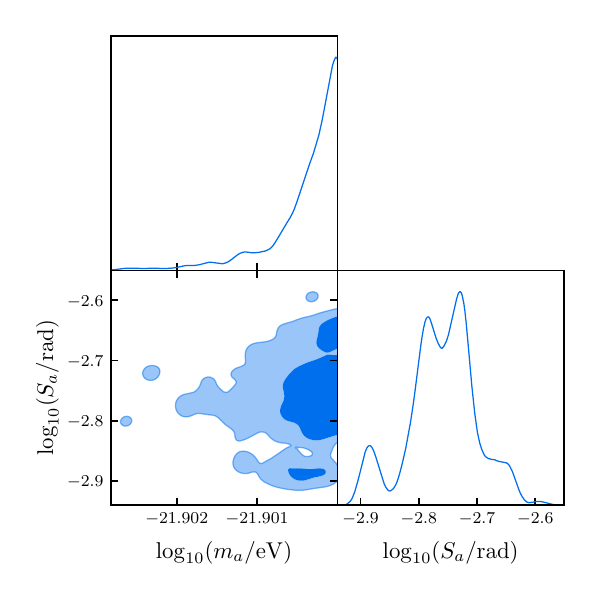}
    \includegraphics[height=0.32\textwidth]{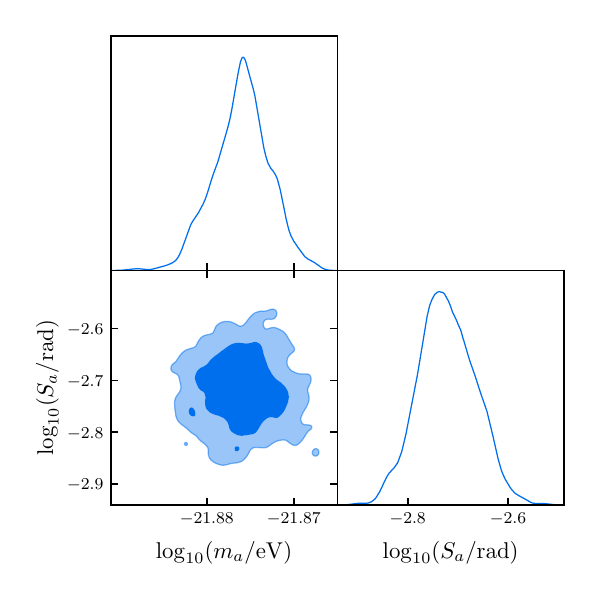}
    \includegraphics[height=0.32\textwidth]{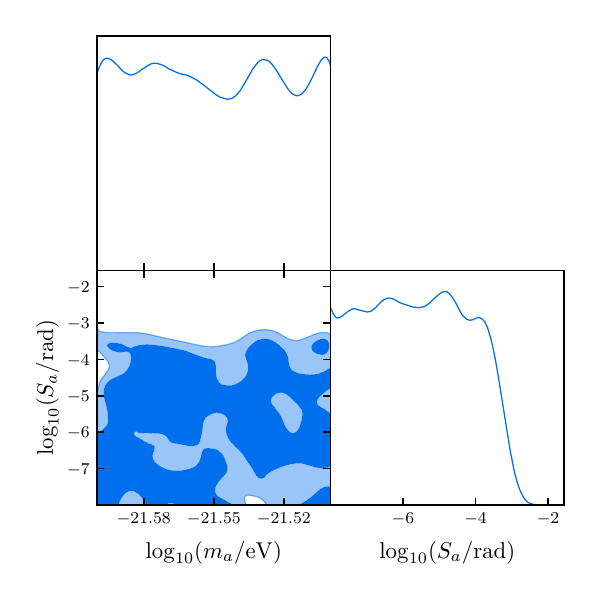} \\
    \caption{
    Posteriors of $m_a$ and $S_a$ near the one-year oscillation period in the data. The TPPA analysis is conducted with  (top row) and without (bottom row) harmonic terms for noise modelling in Eq.~\eqref{eq: IISM}, where the \texttt{spinifex} ionospheric subtraction is applied.  
    The three columns correspond to mass bins with $\mathrm{log}_{10}(m_a/\mathrm{eV})$ in $(-22.0,-21.9)$, $(-21.9,-21.8)$ and $(-21.6,-21.5)$.}
    \label{fig:corner_annual}
\end{figure*}

\begin{figure*}
    \centering
    \includegraphics[width=0.48\linewidth]{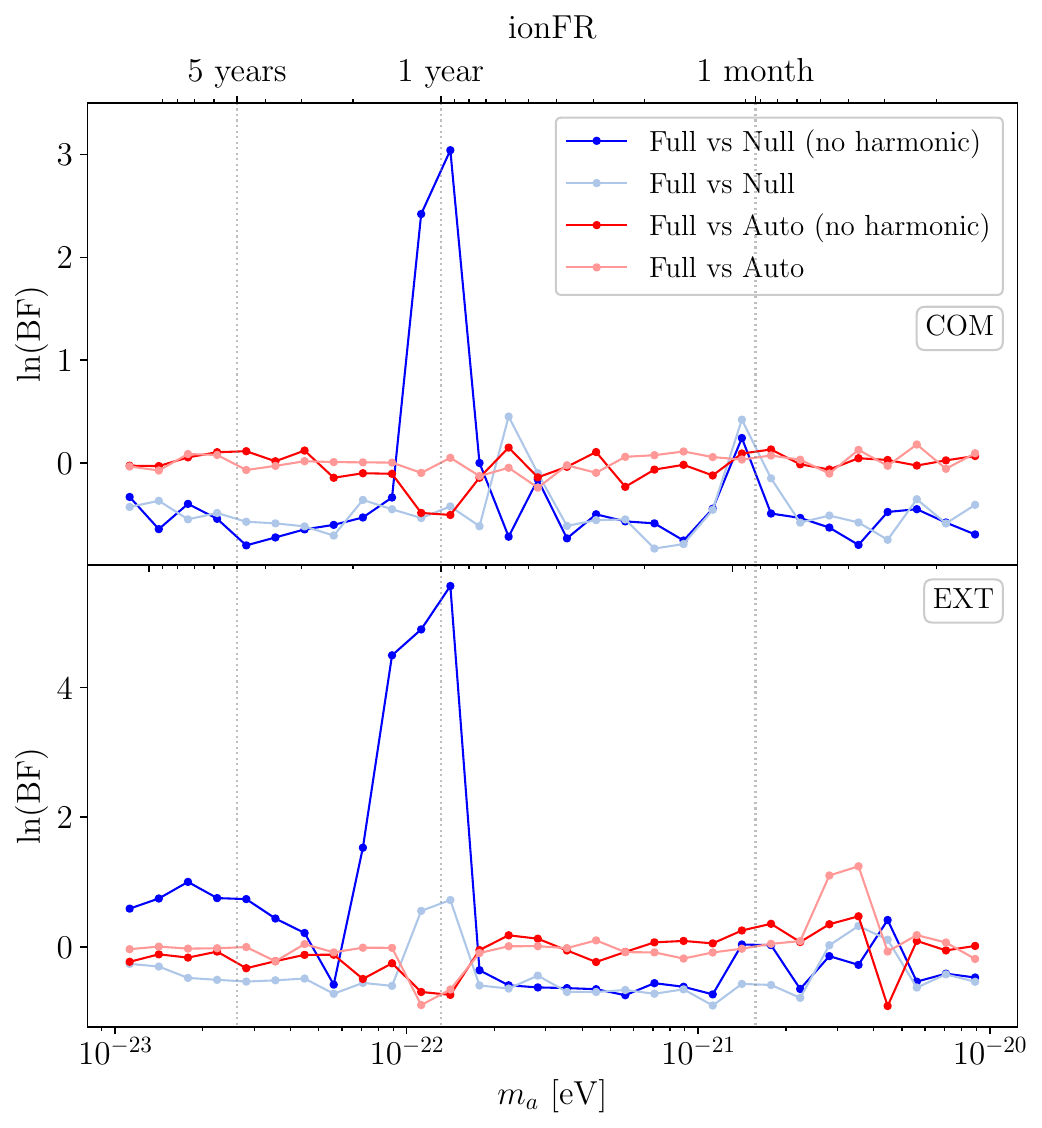}
    \includegraphics[width=0.48\linewidth]{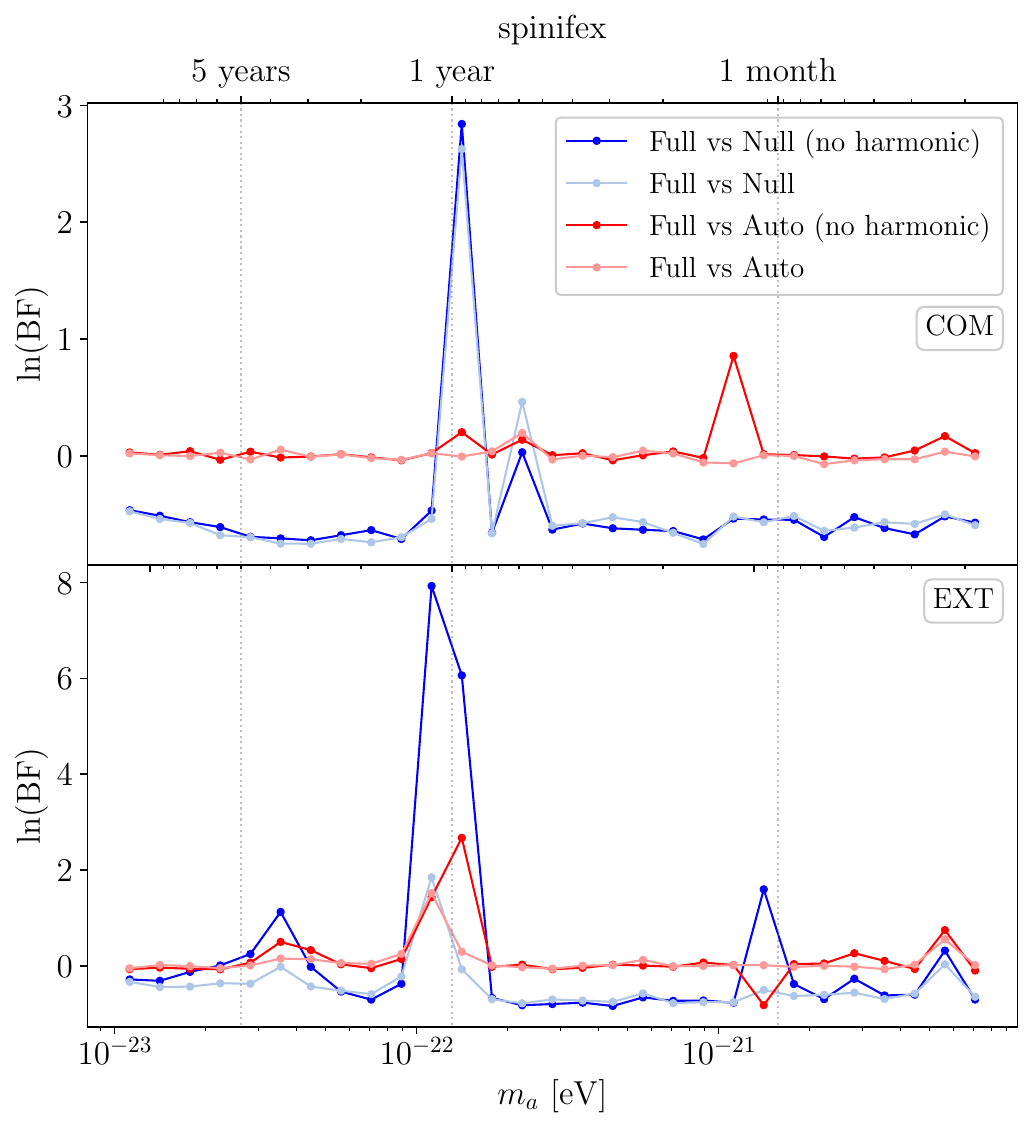}
    \caption{Performance of Bayes factors $\mathrm{BF}_\mathrm{Null}^\mathrm{Full}$ and $\mathrm{BF}_\mathrm{auto}^\mathrm{Full}$ in the TPPA analysis. }
    \label{fig: lnBF ionFR and spinifex}
\end{figure*}



Due to the Earth's orbital motion around the Sun, various seasonal effects can be induced in the pulsar polarization data such as annual variations of the local ionospheric properties and the line-of-sight position of each pulsar in the IISM. If these effects are not properly modelled or subtracted from the data, the exclusion limits could be under estimated at the vicinity of $m_a = 2 \pi/1\,$yr. In this case, the ALDM signal shares the oscillation period of one year with the Earth's orbiting and its strength posterior thus could be trapped to some narrow local parameter region, yielding a bias. The gentle peaks at $m_a = 2 \pi/1\,$yr shown in Fig.~\ref{fig:gagg_main_main} may have indicated this. 

It has been suggested to include harmonic terms, namely $\psi_{p}^{\mathrm{(s)}}\sin\left(2\pi\tilde t \right) + \psi_{p}^{\mathrm{(c)}}\cos\left(2\pi\tilde t\right)$ here, to model such periodic noise effects~\cite{Keith:2012ht,Porayko2024}. In this case, Eqs.~(\ref{eq:Mpsi1}) and (\ref{eq:Mpsi2}) are modified to:
\begin{align}
    &M_p = \left(
    \begin{array}{ccccc}
        1 & \tilde{t}_{p,1} & \tilde{t}_{p,1}^2 & \sin 2\pi\tilde{t}_{p,1} & \cos 2\pi\tilde{t}_{p,1} \\
        1 & \tilde{t}_{p,2} & \tilde{t}_{p,2}^2 & \sin 2\pi\tilde{t}_{p,2} & \cos 2\pi\tilde{t}_{p,2}  \\
        \vdots & \vdots & \vdots & \vdots & \vdots \\
        1 & \tilde{t}_{p,N_p} & \tilde{t}_{p,N_p}^2 & \sin 2\pi\tilde{t}_{p,N_p} & \cos 2\pi\tilde{t}_{p,N_p}
    \end{array} \right)~, \\
    &\psi_p^\transp = \left( \psi_p^{(0)},\psi_p^{(1)},\psi_p^{(2)},\psi_p^{(\mathrm{s})},\psi_p^{(\mathrm{c})} \right) \, .
\end{align}
But, such a treatment is not favored by data as we have demonstrated in~\cite{Sarkis:2025}.

Despite this, we still present some results in this framework here for reference, since they may represent a more conservative consideration: the harmonic terms may cancel the signal's contributions, broadening the posterior of signal strength to larger values and hence weakening the exclusion limits at $m_a = 2 \pi/1\,$yr. Indeed, as shown in the middle panels of  Fig.~\ref{fig:corner_annual}, the posterior of signal strength is narrower for the original noise model and becomes much flatter with the harmonic terms included. For a further comparison, we additionally include the upper limits of TPPAs on the ALDM Chern-Simons coupling in this framework in Fig.~\ref{fig:res of all groups}. The gentle peaks at $T=1\,$yr get visibly sharpened then. 


We also present the $\mathrm{BF}_\mathrm{Null}^\mathrm{Full}$ and $\mathrm{BF}_\mathrm{auto}^\mathrm{Full}$ curves in Fig.~\ref{fig: lnBF ionFR and spinifex} for the noise modelling without and with the harmonic terms. Different from the former case, when the \texttt{ionFR} is applied, no sharp peaks appear at $m_a = 2 \pi/1\,$yr of the $\mathrm{BF}_\mathrm{Null}^\mathrm{Full}$ curves in the latter case, for both COM and EXT TPPAs. This does not imply that these harmonic terms offer a better noise modelling, but indicate that in this case, the inclusion of signals does not improve much data disfavoring. If the \texttt{ionFR} is applied, the outcome is similar for the EXT TPPA analysis. For the COM TPPA analysis, although sharp peaks appear at $m_a = 2 \pi/1\,$yr of $\mathrm{BF}_\mathrm{Null}^\mathrm{Full}$ for both noise modellings, the flatness of the $\mathrm{BF}_\mathrm{Auto}^\mathrm{Full}$ curves indicate null signal in data.

\bibliographystyle{apsrev}
\bibliography{ref}

\end{document}